\documentclass[article,amsmath,amssymb]{revtex4-1}
\usepackage[utf8x]{inputenc}
\usepackage{graphicx}
\usepackage{dcolumn}
\usepackage{bm}
\usepackage{color} 
\usepackage{amsmath}
\usepackage{tensor}

\usepackage{soul} 
\usepackage[normalem]{ulem} 

\begin{document}

\preprint{APS/123-QED}

\title{From generating functions to the geometric Binder cumulant}

\author{Bal\'azs Het\'enyi$^{1,2 *}$}
\affiliation{$^{1}$ Department of Theoretical Physics, Budapest University of Technology and
  Economics, M\H{u}egyetem rkp. 3, H-1111 Budapest, Hungary \\}
\affiliation{$^{2}$Institute for Solid State Physics and Optics, HUN-REN Wigner Research Centre for Physics,  H-1525 Budapest, P. O. Box 49, Hungary\\}

\email{hetenyi.balazs@ttk.bme.hu}

\date{\today}
\begin{abstract}
We present an overview of the role of generating functions in quantum mechanical contexts, mainly in the modern theory of polarization and in the study of quantum phase transitions.  Generating functions enable the derivation of moments and cumulants, quantities which characterize the fluctuations of an underlying probability distribution.  In all of the cases we review, the fluctuations are those of a quantum system.  We show that the original formalism for geometric phases, in which a quantum system is taken around an adiabatic cycle, can be extended to the case when degeneracy points are encountered along the cycle (quasiadiabatic cycles).  The essential tool for this extension is a generalized Bargmann invariant which plays the role of a generating function.  From the cumulants generated this way one can form ratios according to the Binder cumulant scheme in statistical mechanics.  Such geometric Binder cumulants are sensitive to gap closure, as such, they are useful in identifying metal-insulator transitions, localization, and quantum phase transitions.  We present example calculations on simple model systems, whose localization properties are well known, to validate to approach.  We also complement our geometric Binder cumulant calculations with results for the fidelity susceptibility, a quantity directly related to the quantum geometry of the parameter space.
 \end{abstract}

\maketitle

\tableofcontents

\section{Introduction}

The generating function (or characteristic function)~\cite{Lukacs60}, $f(k)$, associated with some probability distribution function, $P(x)$, is simply its Fourier transform.  The moments and cumulants  associated with $P(x)$ are derived from $f(k)$ by taking simple or logarithmic derivatives, respectively, and setting $k$ to zero.  The moments and cumulants are numbers which characterize the probability distribution.  The first moment or cumulant is the average.  Higher order cumulants are independent of the average.  The second cumulant, the square of the variance, characterizes how spread a given distribution is about its mean.  The third cumulant, or skew, gives the extent to which a distribution is asymmetric about its mean.   The fourth cumulant, the kurtosis, characterizes the tails of the distribution.  The wave function in quantum mechanics is a probability amplitude, its modulus square is also a probability distribution.  It follows that the notion of the generating function, as well as moments and cumulants, enter the analysis of quantum systems~\cite{Kubo62,Fulde95}.  This is indeed the case, but generating functions in quantum systems are not always simply the Fourier transform of the probability distribution.  In this review, we explore this question by working through several examples.\\

\begin{figure}[ht]
 \centering
 \includegraphics[width=16cm,keepaspectratio=true]{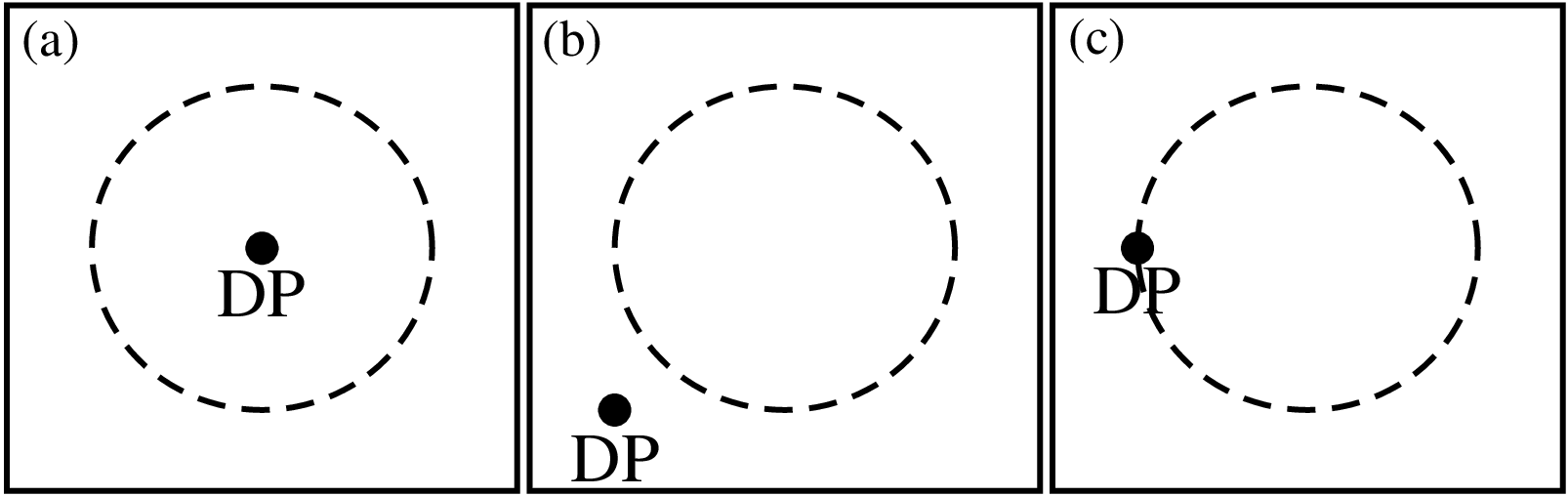}
 \caption{Different scenarios of cyclic curves in a two-dimensional parameter space and a single degenracy point.  The curves are indicated by a dashed line, while the degeneracy point is denoted DP.  Panel (a) shows a curve encircling the degeneracy point, leading to a non-trivial Berry phase.  Panel (b) shows a curve with a degeneracy point not encircled by the curve, leading to a trivial Berry phase ($0$).  Panel (c) shows the curve hitting the degeneracy point.  One question we address is whether it is possible to extend the Berry phase formalism for the case depicted in panel (c).}
 \label{fig:dpcurves}
\end{figure}

A physical quantity unique to quantum mechanical systems is the geometric phase~\cite{Pancharatnam56,LonguetHiggins58,Berry84,Wilczek89,Xiao10} (Berry phase), which arises when the system is taken around an adiabatic cycle in its parameter space.   In its original form the cycle was assumed to be gapped throughout, and the term adiabatic refers to the lack of transitions to excited states which is only exact if the cycle is carried out infinitely slowly.   The process can be understood as the cyclic parallel transport of a Hilbert space vector that represents the state of the quantum system.  Non-trivial values of the geometric phase arise in two ways.   One is if the underlying curvature (Berry curvature) of the parameter space is non-trivial.   The other is if the adiabatic cycle encircles a topological feature.  An example for this second scenario would be a two-dimensional parameter space with a degeneracy point between the energy levels.   If the cycle encircles this degeneracy point, the Berry phase is nonzero.  If the cycle happens to hit the degeneracy point, in which case we can no longer speak of an adiabatic cycle, the geometric phase becomes divergent.  The three possible scenarios are depicted in Fig. \ref{fig:dpcurves}.   We will refer to the case when a cycle crosses an isolated degeneracy point as quasiadiabatic.  One question explored in this review is whether the Berry phase formalism can be extended to account for quasiadiabatic cycles.  We will show that certain quantum generalizations of the notion of a generating function enable such extensions.  \textcolor{black}{We also mention that the adiabatic theorem has been shown to hold for this case~\cite{Kato50}.}\\   

One important manifestation of the geometric phase is in enabling the calculation of the dielectric polarization in crystalline systems~\cite{Zak89,King-Smith93,Resta94,Resta98,Resta99,Aligia99,Ortiz00,Souza00,Resta00,Resta07,Resta10,Spaldin12,Vanderbilt18}.  The average position of charges is simple to calculate in molecules, or systems in which the boundary conditions can be assumed open, but in crystalline systems, where periodic boundary conditions are applied, the position operator becomes ill-defined.   This problem was first treated by King-Smith, Vanderbilt~\cite{King-Smith93} and Resta~\cite{Resta94}, and various derivations of the polarization expression for crystalline systems exist.  These derivations argue that the polarization is, strictly speaking, not a measurable quantity.   Only differences in polarization are experimentally accessible.   Therefore, the key to calculating the polarization is to cast it as a change in polarization expressed as the integrated transient current.  Since the current is related to the phase of the wavefunction, it is not surprising therefore, that the polarization in crystalline systems corresponds to a well-known quantum geometric phase, known as the Zak phase.   The generalization of the modern theory of polarization from band insulators to explicitly correlated ones was given by Resta~\cite{Resta99}.  The variance of the polarization (second cumulant) was derived as an order parameter for metal-insulator transitions by Resta and Sorella~\cite{Resta99}.   This formalism required a filling dependent modification as was shown by Aligia and Ortiz~\cite{Aligia99}.  In the same year a generating function formalism was introduced by Souza, Wilkens, and Martin~\cite{Souza00}.  The modern theory of polarization is now textbook material~\cite{Grosso00,Vanderbilt18} and is covered in several review articles~\cite{Resta00,Resta10,Spaldin12,Resta24}.\\

In statistical mechanics the question of phase transitions~\cite{Cardy96} is a fundamental one.  The analysis of phase transitions often proceeds through scaling and renormalization.  A commonly used approach to identify and characterize phase transition points is the Binder cumulant method~\cite{Binder81a,Binder81b}.  The Binder cumulant is a ratio of statistical cumulants of the order parameter of the system under scrutiny.   Its use stems from the fact that at critical points it can be shown, based on the finite size scaling hypothesis~\cite{Fisher72a,Fisher72b}, to be size independent.  In numerical applications this fact is used to accurately locate and characterize critical points.  In the original formulation it was necessary for the system to exhibit an order parameter, consisting of a sum over local operators, such as the spins on the sites of the system.  There are transitions, however, in which this is not possible.   The metal-insulator transition does not have a local order parameter because, as we know since the seminal work of Kohn~\cite{Kohn64}, a conducting state is defined by the delocalization of the center of mass of all charge carriers in a system, rather than the localization of single charge carriers.  Some quantum systems can also exhibit phase transitions between phases characterized by topological invariants~\cite{Bernevig13,Asboth16,Qi11,Sato17}, rather than usual order parameters, which resist being cast as the sum of local operators.  In such cases the original Binder cumulant formalism is not directly applicable.   As mentioned above, the bulk polarization in a crystalline system is not a usual operator expectation value, but a geometric phase.  The question we address in this context is whether the formalism can be extended to accommodate for the construction of Binder cumulants.   \\

The Berry phase depends crucially on the geometry of the parameter space~\cite{Torma23}.  The parameter space of quantum systems was first characterized systematically by Provost and Vallee~\cite{Provost80}.  This seminal work derived the quantum metric associated with the quantum distance between two quantum states in parameter space.  One can think of the scalar product of two quantum states as a generating function~\cite{Hetenyi23} and derive not only the quantum metric, which corresponds to the second cumulant, but also higher order cumulants (quantum Chistoffel symbol, quantum Riemann curvature tensor).   It is hard to overstate the importance of quantum geometry in assessing the response functions of condensed matter systems~\cite{Torma23,Peotta15,Yu25,Thompson25,Xie25,Verma25,Pellitteri25,Avdoshkin23,Avdoshkin25}.  The Provost-Vallee metric is equivalent to the fidelity susceptibility~\cite{Gu10}, a quantity also used to assess quantum phase transitions~\cite{Sachdev11}.  To complete our work we also present calculations using the fidelity susceptibility. \\

In this work we show that the questions raised above can be addressed by the construction and careful analysis of the relevant generating functions. \\

In the next section (section \ref{sec:basic}) we will give background information relevant for this work.  Since most of our examples are within the modern theory of polarization we will state the problem addressed by this theory, namely, the ill-defined nature of the position operator under periodic boundary conditions.  As further background we define the basic notions of the generating function, moments and cumulants of a probability distribution.   A statistical quantity of interest, the excess kurtosis, which turns out to be closely related to the traditional, order parameter based, Binder cumulant, is given special emphasis.  This section is continued by the description of three different types of geometric phases: the "usual" one (geometric or Berry phase), the open-path Berry phase (also known as Zak phase), and the single point Berry phase.   We derive the discrete form of these geometric phases from their associated Bargmann invariants~\cite{Bargmann64}, and proceed to derive their well-known continuous versions.  The section also introduces the generating function in the context of quantum geometry~\cite{Provost80}.   The quantum metric tensor (or Provost-Vallee metric) becomes the fidelity susceptibility~\cite{Gu10} in the case of a one dimensional parameter space.    \\

In section \ref{sec:MPT} the particular case of periodic probability distributions is analyzed.  We argue that this is the essential element in the construction of the modern theory of polarization, because for periodic probability distributions the extractions of moments and cumulants proceeds through finite difference derivatives of the characteristic function, rather than continuous derivatives.  In this case, even a classical distribution leads to a first cumulant (as well as all other odd order cumulants) described by the {\it phase} of the generating function.  The Zak phase expression follows by substituting the probability distribution of a Slater determinant of Bloch states comprised of one or more filled bands~\cite{Resta98}.   We also show that it is an extension of the Bargmann invariant which plays the role of the generating function in the case of geometric phases.  The geometric phase itself is the first cumulant and higher order cumulants can be generated.  In the continuous limit, higher order cumulants tend to zero, but their ratios, constructed in the same manner as the Binder cumulant, remain finite for quasiadiabatic cycles.  Since the metal-insulator transition and topological quantum phase transitions~\cite{Sachdev11} correspond to gap closures, we propose that the Binder cumulants obtained from extended Bargmann invariants as generating functions be used~\cite{Hetenyi19,Hetenyi22} in the finite size scaling of such transitions.  We refer to these cumulant ratios as geometric Binder cumulants.  \\

\textcolor{black}{A consequence of the use of finite difference derivatives is that the usual equations relating cumulants and moments do not hold between approximate cumulants and approximate moments.  Although, this is not a serious issue on the insulating side, where increasing the order of the approximation makes the two sides of such equations converge to each other, this is not so on the metallic side.  The origin of the problem is that in the cumulant expressions the finite difference derivatives give rise to terms involving the logarithm of the discrete characteristic function, $Z_q$, which go to zero in the conducting phase.  Our geometric Binder cumulant expression avoids this pitfall, because it is a ratio of centered moments rather than cumulants, which do not have $\mbox{log} Z_q$ terms, only terms depending on $|Z_q|$.  This ansatz preserves the correct scaling of each relevant quantity with system size and is applicable on both sides of the metal-insulator transition.} \\

Example calculations to validate the formalism are presented in section \ref{sec:xmpls}.  An essential, but easy, example is the simple tight-binding model at half filling, which is the simplest model for a conductor on a lattice.  This is obviously a gapless system.  The polarization has easily accessible distributions with known excess kurtoses (geometric Binder cumulants).   We calculate the Binder cumulant of the polarization for this system, and we find that it takes a value of $0.4$, which is exactly the known excess kurtosis of the flat polarization distribution.  We then consider the same model under periodic boundary conditions and calculate the analogous quantity, the geometric Binder cumulant.  We find that if the order of the approximation is increased the value of $0.4$ is approached.   We also consider a periodic tight-binding system with a degenerate groundstate, whose polarization distribution can be shown to be a raised cosine distribution.  The geometric Binder cumulant, in the limit of large order finite difference approximation, reproduces the value known from the literature for this distribution as well.  We believe that these results for simple text cases validate our method.  For completeness an example of a topological phase transition is also considered, the Su-Schrieffer-Heeger model~\cite{Su79}.   We also present calculations for the Aubry-Andr\'{e} model ~\cite{Aubry80}.  We compare the commonly used scheme of Resta-Sorella (and its extended version by considering higher order finite difference approximations), and a slightly different approach advocated here.   For the variance of the polarization the two approaches converge to the same value in the insulating phase of the model, but they deviate in the delocalized phase.  Our modified approach reproduces known size scaling in both the delocalized and localized cases.  And finally, in this section, we also calculate the fidelity susceptibility in the Aubry-Andr\'{e} model.  Our results, again, agree with recent results~\cite{Cookmeyer20,Hetenyi25} on this model.  To give a complete picture of the Aubry-Andr\'{e} model we also present the number theoretical rationalization of the localization transition.\\

In section \ref{sec:cnclsn} we conclude our work. \\

\section{Basic ingredients}

\label{sec:basic}

\subsection{The problem of the polarization in crystalline systems}

When the polarization is calculated for any condensed matter system, the usual starting point is the Born-Oppenheimer approximation which assumes that the nuclei of the system are fixed, and that it is only the charge carrying electrons which  move~\cite{Grosso00,Vanderbilt18}.  The dielectric polarization of such systems can be written as
\begin{equation}
{\bf P} = {\bf P}_{el} + {\bf P}_{nuc}, 
\end{equation}
where the electronic contribution is given by,
\begin{equation}
{\bf P}_{el} = -\frac{e N_e}{V}\int d{\bf r} \rho({\bf r}) {\bf r},  
\end{equation}
and the nuclear one by,
\begin{equation}
{\bf P}_{nuc} = \frac{e}{V}\sum_{I=1}^N Z_I {\bf R}_I.  
\end{equation}
The electronic contribution is an expectation value over the position operator.  The polarization, in the case of a system with open boundary conditions, is an average dipole moment. \\

In crystalline systems the bulk polarization is calculated~\cite{Resta00,Resta10,Spaldin12,Resta24} under periodic boundary conditions.  The nuclei form an infinite regular lattice, so the potential due to them is periodic.   For a one-dimensional system we can take this periodicity to be specified by the lattice parameter, $l$.     The electrons, on the other hand, can delocalize over many unit cells, even in insulators.   For the electronic subsystem, consisting of $N_e$ electrons, the wavefunction satisfies,
\begin{equation}
\label{eqn:Psi}
\Psi(x_1,...,x_j,...,x_{N_e}) = \Psi(x_1,...,x_j+L,...,x_{N_e}); \forall j ,
\end{equation}
where $L = N_c l$, $N_c \in \mathbb{Z}$.  In this case the position operator becomes ill-defined, meaning that the expression given above for ${\bf P}_{el}$ can not be used to calculate the electronic contribution of the polarization.  This problem has been a well-known problem since the 1990s and its solution~\cite{King-Smith93,Resta94} is to replace the expectation value of the position by a geometric phase.  \\

\subsection{Generating functions, moments, cumulants}

Moments and cumulants are numbers which characterize probability distributions~\cite{Lukacs60}.  In every day life, a probability distribution underlying an event or process is often unknown, but an idea can be gained about it, by specifying its cumulants and related quantities.  For example, one might wish to visit Nairobi in September, and wonder about the weather there, to know whether to dress warm, cold, to take an umbrella, etc.  To assess one's prospects, one may look up data, such as the average temperature, as well as the variance or the same info about other quantities, such as precipitation, etc.  The average is the first cumulant (or first moment), whereas the variance is the second cumulant.  While the true probability distribution in this case is not known, from the data collected over time the average and the variance is usually available.\\

Given a probability distribution function, $P(x)$, satisfying,
\begin{equation}
P(x) \geq 0, \int _{-\infty}^\infty d x P(x) = 1.
\end{equation}
A well-behaved probability distribution goes to zero at $x = \pm \infty$.  The moments and cumulants of $P(x)$ can be obtained from the cumulant generating (or characteristic) function~\cite{Lukacs60},
\begin{equation}
\label{eqn:Px}
f(k) = \int_{-\infty}^\infty d x P(x) e^{i k x}.
\end{equation}
The definition of the $n$th moment is,
\begin{equation}
\label{eqn:M_n}
M_n = \frac{1}{i^n}  \left.  \frac{\partial^n f(k)}{ \partial k^n}  \right|_{k=0},
\end{equation}
whereas the definition of the $n$th cumulant is,
\begin{equation}
\label{eqn:C_n}
C_n = \frac{1}{i^n} \left. \frac{\partial^n \ln f(k)}{ \partial k^n} \right|_{k=0}.
\end{equation}
Cumulants can be written in terms of moments, and vice versa, the first four $C_n$ can be written as,
\begin{eqnarray}
\label{eqn:C_M}
C_1 &=& M_1, \\ \nonumber
C_2 &=& M_2 - M_1^2, \\ \nonumber
C_3 &=& M_3 - 3 M_2 M_1 + 2 M_1^3, \\ \nonumber
C_4 &=& M_4 - 4 M_3 M_1 - 3 M_2^2 + 12 M_2 M_1^2 - 6 M_1^4.
\end{eqnarray}
One can also shift the probability distribution so that it is centered, i.e., its average is zero.  If the $M_1$ is the first moment of the distribution $P(x)$, then $P(x+M_1)$ averages to zero.  The corresponding characteristic function is modified as
\begin{equation}
f(k) \rightarrow f(k) e^{-i k M_1}.
\end{equation}
Generating moments and cumulants follow Eqs. (\ref{eqn:M_n}) and (\ref{eqn:C_n}), respectively, using the modified characteristic function.
In this case, the relations between cumulants and moments become,
\begin{eqnarray}
\label{eqn:C_M_centered}
C_1 &=& 0, \\ \nonumber
C_2 &=& M_2, \\ \nonumber
C_3 &=& M_3, \\ \nonumber
C_4 &=& M_4 - 3 M_2^2.
\end{eqnarray}
\\
Probability distributions are usually characterized by their cumulants, or quantities constructed from them.  The average and the variance are common, but higher order cumulants are also used.  The skewness measures the asymmetry of a random distribution about its mean, and it is given by,
\begin{equation}
\tilde{\mu}_3 = \frac{C_3}{C_2^\frac{3}{2}}.
\end{equation}
A distribution with $\tilde{\mu}_3 = 0$ is symmetric about its mean, the sign of $\tilde{\mu}$ gives the side of the mean on which a distribution has a sharper peak.  The excess kurtosis measures how "tailed" a distribution is, and it is given by,
\begin{equation}
\label{eqn:XSkurtosis}
\tilde{\kappa}_4 = \frac{C_4}{C_2^2},
\end{equation}
or, for centered moments,
\begin{equation}
\tilde{\kappa}_4 = \frac{M_4}{M_2^2} - 3.
\end{equation}
For a Gaussian distribution, $\tilde{\kappa}_4 = 0$.  Distributions with $\tilde{\kappa}_4 < 0$($\tilde{\kappa}_4 >0$) are referred to as platykurtic(leptokurtic).  A platykurtic distribution produces less outliers than a Gaussian.   Below we will make use of two well known results: the excess kurtosis of the flat distribution is $-1.2$, and of the raised cosine distribution, it is 
\begin{equation}
\label{eqn:RC}
\tilde{\kappa}_4 = \frac{6}{5} \frac{90 - \pi^4}{(\pi^2 - 6)^2} \approx -.593762.....
\end{equation}
\\

\subsection{The excess kurtosis and the Binder cumulant}

The excess kurtosis has found use in computational statistical physics.  The quantity known as the Binder cumulant~\cite{Binder81a,Binder81b} is used in Monte Carlo simulations of many-body systems to locate critical points.   The precise determination of critical points is difficult through the calculation of the quantities available to the experimentalist.  For example, in experiment, the divergence in the susceptibility locates the critical point exactly, but in a Monte Carlo simulation the system sizes are too small: there is no real divergence, and there is usually a shifting and/or smearing in the calculated critical point.  This is the problem solved by the Binder cumulant, which is a ratio of moments or cumulants of equal "overall" exponents.   \textcolor{black}{The justification for this originates~\cite{Fisher72a,Fisher72b} in the theory of scaling and renormalization near critical points}. By equal overall exponents, I mean a ratio of cumulants, such as $\frac{C_m}{(C_n)^l}$ such that $m = n \times l$.  The most commonly used Binder cumulant is the excess kurtosis of the order parameter (times $-1/3$),  Its form is
\begin{equation}
\label{eqn:U4}
U_4 = 1 - \frac{M_4}{3 M_2^2},
\end{equation}
and the variable averaged in $M_4$ or $M_2$ is the order parameter of the system.  For example, in the Ising model, it is the magnetization.  How does $U_4$ locate critical points?  If one is to locate the critical point of the Ising model, one carries out temperature scans for different system sizes.  It can be estabilshed from scaling theory (in particular finite size scaling) that the Binder cumulant is size independent at the critical point, therefore, the finite temperature value at which the calculated curves for different system sizes cross is the critical temperature.  

\subsection{Berry phases}

In this section we first construct three different types of Berry phases: the quantity which itself is usually called the Berry phase~\cite{Berry84}, the open path Berry phase~\cite{Zak89} (or Zak phase), and the single point Berry phase~\cite{Resta98}.   The first two of these have a discrete and a continuous version.  We derive their discrete versions from a quantity known as the Bargmann invariant and then use these to derive their continuous versions.   Extensions of the  Bargmann invariants~\cite{Bargmann64} defined here will play the role of the generating functions in different cases. \\

\subsubsection{The Berry phase}

Consider a quantum system whose Hamiltonian, $H(\xi)$, depends on a set of parameters denoted by vector, $\xi$.  The vector space in which $\xi$ lives is assumed to be finite dimensional.   We also assume that the normalized ground state wavefunction, $| \Psi _0(\xi) \rangle$, satisfying,
\begin{equation}
H(\xi) | \Psi_0 (\xi) \rangle = E_0(\xi) | \Psi_0(\xi) \rangle,
\end{equation}
is known for some region of the parameter space $\xi$.   Furthermore, we consider a set of points, $\xi_0,...,\xi_{M-1}$, in the parameter space.  We further assume that the first excited state is separated from the ground state by an energy gap for all $\xi_J$.  \\

We consider the Bargmann invariant, the product,
\begin{equation}
\label{eqn:Gamma_1}
\Gamma_1 = \prod_{J=0}^{M-1} \langle \Psi_0 (\xi_J) | \Psi_0(\xi_{J+1})\rangle.
\end{equation}
We assume that the product is cyclic, meaning $| \Psi_0 (\xi_M) \rangle = | \Psi_0 (\xi_0) \rangle$.  This product is a {\it physically well-defined quantity} because the arbitrary phase of each $|\Psi(\xi_J)\rangle$ is canceled by its dual vector $\langle \Psi (\xi_J) |$.  From the Bargmann invariant, the discrete Berry phase is defined as,
\begin{equation}
\label{eqn:g1_lnG1}
\gamma_1 = \mbox{Im} \log \Gamma_1.
\end{equation}
\textcolor{black}{$\gamma_1$ is only defined modulo $2\pi$}. To arrive at the continuous version, one assumes that the parameter set  $\xi_0,...,\xi_{M-1}$ is distributed in order along some closed continuous curve.  Assuming the distance between neighboring points is small, one can then carry out a Taylor expansion up to first order in each scalar product as,
\begin{eqnarray}
\label{eqn:ggg1}
\nonumber
\gamma_1 &=& \mbox{Im} \sum_{J=1}^M \log \langle \Psi (\xi_J)| \left\{ | \Psi(\xi_J) \rangle + \Delta \xi_J \cdot \nabla_\xi | \Psi (\xi_J) \rangle \right\}, \\
 &=& \mbox{Im} \sum_{J=0}^M  \Delta \xi_J \cdot \langle \Psi (\xi_J)| \nabla_\xi | \Psi (\xi_J) \rangle 
\end{eqnarray}
where $\Delta \xi_J = \xi_{J+1} - \xi_J$.   Taking the two limits, $M \rightarrow \infty$, and $\Delta \xi_J \rightarrow 0$, simultaneously results in the cyclic integral,
\begin{equation}
\label{eqn:g1_Berry}
\gamma_1 = \mbox{Im} \oint d \xi \cdot \langle \Psi_0 (\xi)| \nabla_\xi | \Psi_0(\xi)\rangle,
\end{equation}
commonly known as the Berry phase.   \textcolor{black}{By considering a gauge transformation of the type,}
\begin{equation}
| \Psi_0(\xi) \rangle \rightarrow \exp(i \alpha(\xi)) | \Psi_0(\xi) \rangle, 
\end{equation}
\textcolor{black}{the transformed $\gamma_1$ differs by $\oint d \xi \cdot \nabla \alpha(\xi)$.  Since $\alpha(\xi)$ is the phase of the wavefunction, even for a single valued wave function, $\alpha(\xi)$ can take multiple values at the same point $\xi$.  The the circuit integral itself,  $\gamma_1$, is also only defined modulo $2 \pi$.}   Via Stokes theorem $\gamma_1$ can be written,
\begin{equation}
\label{eqn:BC}
\gamma_1 = \int \int  \langle  \nabla_\xi \Psi_0 (\xi)| \times | \nabla_\xi  \Psi_0(\xi)\rangle \cdot d \sigma.
\end{equation}
The quantity $\langle  \nabla_\xi \Psi_0 (\xi)| \times | \nabla_\xi  \Psi_0(\xi)\rangle$ is known as the Berry curvature. \\

The circuit integral, $\gamma_1$, being a physically well-defined quantity, has an extensive literature.  As mentioned before, one interesting example occurs when the space $\xi$ is two dimensional, and a degeneracy point exists in that space.  In this case, the circuit integral defining $\gamma_1$ can encircle the degeneracy point leading to  $\gamma_1$ taking a nonzero value.  If the wavefunction is restricted to be real (possible in some cases, for example, for the ground state of a time-reversal symmetric system), $\gamma_1=\pi$, while if the circuit does not encircle the degeneracy point (trivial case), $\gamma_1=0$.\\

\subsubsection{The open path Berry phase (Zak phase)}

It was first found by Zak that a physically well-defined quantity similar to the Berry phase can be constructed for open paths too, provided that the endpoints are connected by a symmetry operation.\\

We consider a sequence, $\xi_1,...,\xi_M$.  Let there be a symmetry relation, denoted by the operator $\hat{S}$, between the ground state wave functions of the Hamiltonian $H({\bf \xi})$ at the endpoints of the sequence, $\xi_M$ and $\xi_0$:
\begin{equation}
| \Psi(\xi_{M}) \rangle = \hat{S} | \Psi(\xi_0) \rangle. 
\end{equation}
In this case, one can define a Bargmann invariant of the form
\begin{equation}
\label{eqn:Gamma_1_S}
\Gamma_1^{(S)} = \langle \Psi_0 (\xi_0) | \Psi_0(\xi_1)\rangle  \langle \Psi_0 (\xi_1) | \Psi_0(\xi_2)\rangle...\langle \Psi_0 (\xi_{M-1}) | \Psi_0(\xi_M)\rangle.
\end{equation}
The ket vector $|\Psi_0(\xi_M)\rangle$ can be "pulled back" to $|\Psi_0(\xi_0)\rangle$ by use of the symmetry operator $\hat{S}$.  Applying it to Eq. (\ref{eqn:Gamma_1_S}) results in,
\begin{equation}
\label{eqn:Gamma_1_S2}
\Gamma_1^{(S)} = \langle \Psi_0 (\xi_0) | \Psi_0(\xi_1)\rangle  \langle \Psi_0 (\xi_1) | \Psi_0(\xi_2)\rangle... \langle \Psi_0 (\xi_{M-1}) | \hat{S}| \Psi_0(\xi_0)\rangle.
\end{equation}
$\Gamma_1^{(S)}$ is still independent of arbitrary phases because for each $| \Psi(\xi_J) \rangle$ the dual $\langle \Psi(\xi_J)|$ also appears in the product.\\

As an example, we derive the the Berry-Zak phase corresponding to dielectric polarization in the modern theory of polarization.   The parameter space in this case is the Brillouin zone (the parameter itself being the crystal momentum), the relevant wavefunctions are the Bloch states of one or more bands, and the symmetry operator is the total momentum shift (or twist) operator.  For simplicity we consider a one dimensional system and we will derive the Berry-Zak phase for a single band.  For a one-dimensional system, we can consider the crystal momentum in the first Brillouin zone, $-\frac{\pi}{l} < k \leq \frac{\pi}{l}$, and write the Bloch states as,
\begin{equation}
\tilde{\Psi}_k (x) = e^{i k x} u_k(x),
\end{equation}
where $u_k(x)$ is a periodic Bloch function, satisfying $u_k(x+L)=u_k(x)$.   \textcolor{black}{We work in the "periodic gauge" in which the phases of the periodic Bloch functions are chosen so that $\tilde{\Psi}_k(x) = \tilde{\Psi}_{k + \frac{2 \pi}{l}}(x)$}. We will construct a discrete geometric phase.  We consider the points $k_m = \frac{2 \pi}{L}m$ where $L = N_c l$ and $m = 0,...,N_c$, which constitutes an open path.   We write $\Gamma_1$ according to Eq. (\ref{eqn:Gamma_q}) as
\begin{equation}
\Gamma_1^{(Z)} = \langle u_{k_0} | u_{k_1} \rangle  \langle u_{k_1} | u_{k_2}\rangle...\langle u_{k_{N_c-1}} | u_{k_{N_c}}\rangle,
\end{equation}
Since, in the periodic gauge, 
\begin{equation}
u_{k+\frac{2 \pi}{l}}(x) = e^{i \frac{2 \pi x}{l}} u_k(x),
\end{equation}
we can define the symmetry operator,
\begin{equation}
\hat{S} = e^{- i \frac{2 \pi x}{l}},
\end{equation}
and rewrite the phase $\Gamma_1^{(Z)}$ as,
\begin{equation}
\label{eqn:Gamma_1_Zak}
\Gamma_1^{(Z)} = \langle u_{k_0} | u_{k_1} \rangle  \langle u_{k_1} | u_{k_2}\rangle...\langle u_{k_{N_c-1}} | \hat{S}| u_{k_0}\rangle,
\end{equation}
\

The continuous Berry-Zak phase can be derived as before,
\begin{equation}
\label{eqn:gamma_Z}
\gamma_Z = \mbox{Im} \int_{-\frac{\pi}{l}}^{\frac{\pi}{l}} d k \langle u_k | \frac{\partial}{ \partial k}| u_k \rangle.
\end{equation}
\textcolor{black}{Like the Berry phase derived earlier, $\gamma_Z$ is also only defined modulo $2 \pi$.  In the modern theory of polarization~\cite{King-Smith93,Resta94,Resta98,Resta00,Resta07,Spaldin12,Vanderbilt18} the indeterminacy in the Zak phase (which corresponds to an indeterminacy in the bulk polarization) is interpreted as the indeterminacy one expects due to the presence of boundary charges.}

\subsubsection{The single point Berry phase}

The single point Berry phase is simply the phase of the expectation value of some unitary operator, $\hat{U}$.   The wavefunction is only needed at a single point in parameter space, we can simply denote it $|\Psi \rangle$
\begin{equation}
\gamma = \mbox{Im} \log \langle \Psi | \hat{U} | \Psi \rangle.
\end{equation}
\textcolor{black}{$\gamma$ is defined up to modulo $2 \pi$.}  An example of a single point Berry phase was introduced by Resta to describe the polarization in many-body systems with periodic boundary conditions.  This phase takes the form,
\begin{equation}
\gamma_R = \mbox{Im} \log \langle \Psi | \exp \left( i \frac{2 \pi}{L} \hat{X} \right) | \Psi \rangle.
\end{equation}
The expectation value of the total position, needed to express the dielectric polarization, can be written,
\begin{equation}
\langle X \rangle = \frac{L}{2 \pi} \gamma_R.
\end{equation}
From $\gamma_R$, the Zak phase (Eq. (\ref{eqn:gamma_Z})) can be recovered in the non-interacting case.  If the many-body state $| \Psi \rangle$ is a Slater determinant of single particle orbitals filling a band, then $\gamma_R$ can be shown~\cite{Resta98,Resta99} to be the Zak phase, $\gamma_Z$ of that particular band.\\

\subsection{The generating function in quantum geometry}

We mention one last example of a generating function, namely, the one used in quantum geometry~\cite{Provost80}.   Given, as before, a parametrized quantum state, $| \Psi (\xi) \rangle $, the generating function~\cite{Hetenyi23} in this case is simply the scalar product, $\langle \Psi(\xi) | \Psi(\xi') \rangle$.  Cumulants of order $m+n$ can be generated as,
\begin{equation}
C_{m+n}(k_1,...,k_m;l_1,...,l_n) = \left(-i \frac{\partial}{\partial \xi_{k_1}}\right)...\left(-i \frac{\partial}{\partial \xi_{k_m}}\right)\left(i \frac{\partial}{\partial \xi'_{l_1}}\right)...\left(-i \frac{\partial}{\partial \xi'_{l_n}}\right) \log \left. \langle \Psi (\xi) | \Psi (\xi') \rangle \right|_{\xi' = \xi}.
\end{equation}
Perhaps, most commonly analyzed is the second cumulant,
\begin{equation}
C_2(k;l) =  \left.  \frac{ \partial^2 \log \langle \Psi (\xi) | \Psi (\xi') \rangle}{\partial \xi_k \partial \xi'_l} \right|_{\xi' = \xi},
\end{equation}
which has a real and an imaginary part.  The real part takes the form,
\begin{equation}
g_{kl} = \langle \partial_{\xi_k} \Psi (\xi) | \partial_{\xi_l} \Psi(\xi) \rangle -  \langle \partial_{\xi_k} \Psi (\xi) | \Psi(\xi) \rangle \langle \Psi(\xi) \partial_{\xi_l} \Psi(\xi) \rangle,
\end{equation}
and is known as the Provost-Vallee metric.  The imaginary part is the Berry curvature, which was given in Eq. (\ref{eqn:BC}).  In principle the Provost-Vallee metric exists even for a one-dimensional parameter space.  In this case, it is also known as the fidelity susceptibility, used in the analysis of quantum phase transitions.  Driving the fidelity susceptibility across phase transitions can be used to identify quantum critical points.\\

\section{Generating functions in crystalline systems}

\label{sec:MPT}

In this section we develop generating functions in the context of the modern theory of polarization.  The first step is to consider how a generating function changes if the probability distribution from which it is derived is periodic.  What changes is that the argument of the generating function is only available at a discrete set of points, not the entire real line.  This leads to an important change in the formalism: the derivatives become approximate, finite difference derivatives.  These two ingredients suffice to define moments and cumulants of the polarization in interacting crystalline systems.   We then introduce the gauge invariant cumulants introduced by Souza, Wilkens, and Martin~\cite{Souza00}.  While the cumulants developed here are not the same as the SWM gauge invariant cumulants, they need to be mentioned, not only because they were an important development in the MPT formalism, but also, because they lead to $\mathcal{O}(1)$ numbers in the insulating state, and we will use this fact to show that the geometric Binder cumulant is zero there.  \\

It is then shown that an extension of the Bargmann invariant can be used in place of a generating function for the Berry phase.  The Berry phase is a first moment or cumulant, but higher order moments and cumulants can also be derived in the discrete case.   When this approach is modified to accommodate for the case of the open path Berry phase, the cumulants of the polarization result.  The Binder cumulant for polarization can be constructed from the cumulants obtained this way.\\

\subsection{Periodic probability distributions}

Starting with a usual probability distribution as defined in Eq. (\ref{eqn:Px}), we can "periodize" it via
\begin{equation}
P_l(x) = \sum_{w = - \infty}^\infty P(x + w l),
\end{equation}
where $l$ denotes the periodicity length.  We note that in a crystalline quantum system the sum of all the modulus squared of all Bloch functions of a given band lead to a periodic function, similar to $P_l(x)$, while the modulus squared of a Wannier function associated with a given band behaves as $P(x)$.  \\

\textcolor{black}{ The continuous Fourier transform can not be applied to  $P_l(x)$, instead, it can be written as a Fourier series is over a discrete set of $k$ values.}  The characteristic function takes the form,
\begin{equation}
f_q = \int_0^ l d x P_l(x) e^{i k_q x}
\end{equation}
where
\begin{equation}
k_q = \frac{2 \pi}{l} q; q \in \mathbb{Z}.
\end{equation}  
\\

Since the characteristic function is not available on the entire real axis, but, instead, only on a discrete set of points, one can not calculate the moments or the cumulants via Eqs. (\ref{eqn:M_n}) and (\ref{eqn:C_n}), because one can not take continuous derivatives.  Instead, one has to resort to finite difference derivatives, which are approximations to the derivative.  We write,  
\begin{eqnarray}
\label{eqn:MC_fd}
M_n &=& \left( \frac{l}{2 \pi i} \right)^n \left. \frac{\delta^n f_q}{ \delta q^n}  \right|_{q=0}, \\ \nonumber
C_n &=& \left( \frac{l}{2 \pi i} \right)^n \left. \frac{\delta^n \mbox{Log} f_q }{ \delta q^n}  \right|_{q=0},
\end{eqnarray}
where $\frac{\delta}{\delta q}$ denotes a finite difference derivative with respect to $q$.    \textcolor{black}{In $C_n$ we restricted the logarithm of $f_q$ to be on the principal branch, $(\pi,\pi]$.  The reason for this is because the derivative is to be evaluated at $q=0$ and when the limit of the finite difference derivatives is taken the points closest to zero will give the correct value for the derivative.   We also introduced the notation $\mbox{Log}$ for complex logarithm in which the phase is taken to be on the principal branch.   As mentioned above, in the definition of the Zak phase (and crystalline polarization), the modern polarization theory allows for the multivaluedness of the relevant phase, and interprets this indeterminacy as due to the possible presence of surface charges.   The presence of surface charges shifts the first cumulant (the mean), but higher order cumulants do not depend on the mean.  In this sense, our restriction for higher order odd cumulants is justified.  Our restriction is also justified for even cumulants which are magnitudes, rather than phases, and for which such an indeterminacy does not exist. }\\

Due to the use of finite difference derivatives, Eqs. (\ref{eqn:C_M}) and (\ref{eqn:C_M_centered}) are no longer valid.   It is possible to take an integer number of unit cells as the fundamental periodicity length (supercells), $L= N_c l$, and obtain the related moments and cumulants.    The large $N_c$ limits,
\begin{eqnarray}
\label{eqn:MC_fd}
M_n &=& \lim_{N_c \rightarrow \infty} \left( \frac{N_c l}{2 \pi i} \right)^n \left. \frac{\delta^n f_q}{ \delta q^n}  \right|_{q=0}, \\ \nonumber
C_n &=& \lim_{N_c \rightarrow \infty} \left( \frac{N_c l}{2 \pi i} \right)^n \left. \frac{\delta^n  \mbox{Log} f_q}{ \delta q^n} \right|_{q=0},
\end{eqnarray}
correspond to the thermodynamic limit of $M_n$ and $C_n$.  For insulating (gapped) systems, the relations  Eqs. (\ref{eqn:C_M}) and (\ref{eqn:C_M_centered}) are restored.  This is not true for conducting or gapless systems.  \\

The first four lowest order finite difference derivatives have the form,
\begin{eqnarray}
\left. \frac{\delta}{\delta q} f_q \right |_{q=0} &=& \frac{f_1 - f_{-1}}{2}, \\ \nonumber
\left. \frac{\delta^2}{\delta q^2} f_q \right |_{q=0} &=& 2(f_1 + f_{-1} - 2 f_0), \\ \nonumber
\left. \frac{\delta^3}{\delta q^3} f_q \right |_{q=0} &=& \frac{ f_2 - 2 f_1 + 2 f_{-1} - f_2}{2}, \\ \nonumber
\left. \frac{\delta^4}{\delta q^4} f_q \right |_{q=0} &=& (f_2 - 2 f_1 + 2 f_0 -  2 f_{-1} + f_{-2}).
\end{eqnarray}
which are correct up to order $\mathcal{O}(L^{-2})$.  For the first four cumulants we obtain,
\begin{eqnarray}
\label{eqn:C_periodic}
C_1 &=& \frac{L}{2 \pi} \mbox{Im} \mbox{Log} f_1, \\ \nonumber
C_2 &=& -\frac{L^2}{2 \pi^2} \mbox{Re} \mbox{Log} f_1 \\ \nonumber
C_3 &=& -\frac{L^3}{8 \pi^3} (\mbox{Im}\mbox{Log} f_2 - 2 \mbox{Im} \mbox{Log} f_1), \\ \nonumber
C_4 &=& \frac{L^4}{8 \pi^4} ( \mbox{Re}\mbox{Log} f_2 - 4 \mbox{Re} \mbox{Log} f_1).
\end{eqnarray} 
In deriving these expressions we used the fact that $f_0 = 1$ and that $f_q = f_{-q}^*$.  It is to be noted that odd cumulants of the discrete characteristic function correspond to sums of phases, the even ones to sums of magnitudes.  If one considers higher order finite difference approximations this state of affairs is unchanged: odd cumulants correspond to sums of phases, even ones to sums of magnitudes.  On the other hand, Eq. (\ref{eqn:C_periodic}) is expected to hold for $N_c \rightarrow \infty$.  \\

\subsection{The modern polarization theory for crystalline many-body systems}

The probability distribution obtained from a wave function obeying periodic boundary conditions is periodic, moreover, the distribution of the total position will share the same periodicity.   Using the form of the wave function in Eq. (\ref{eqn:Psi}), the probability distribution can be written,
\begin{equation}
P_\Psi(x) = \int...\int d x_1...d x_{N_e} | \Psi(x_1,...,x_{N_e})|^2 \delta\left( x - \sum_j^{N_e} x_j \right),
\end{equation}
and obviously, $P_\Psi(x + l) = P_\Psi(x)$.  The characteristic function in this case, usually labeled $Z_q$, is a single point Berry phase, and takes the form,
\begin{equation}
Z_q = \langle \Psi | \exp \left( i \frac{2 \pi q}{L} \hat{X} \right)| \Psi \rangle,
\end{equation}
where $\hat{X} = \sum_{j = 1}^{N_e} \hat{x}_j$ denotes the total position operator.  $Z_q$ is entirely analogous to a discrete characteristic function, $f_q$, derived in the previous section: moments and cumulants can be accessed through Eqs. (\ref{eqn:MC_fd}) and (\ref{eqn:C_periodic}). \\

Stated in the simplest terms the modern polarization theory replaces the total position expectation value of the electronic degrees of freedom with the first cumulant of the distribution $P_\Psi(x)$, using finite difference approximations to the derivatives appearing in Eqs. (\ref{eqn:MC_fd}).   Since, in this case, the finite difference derivative is applied to $\ln Z_q$, we will refer to this approach as the finite difference logarithmic derivative (FDLD) approach.  The lowest order FDLD based approach for $C_1$ and $C_2$ correspond to the many-body polarization given by Resta and the Resta-Sorella variance, respectively.   They take the forms,
\begin{eqnarray}
\label{eqn:Resta_Sorella}
C_1 &=& \frac{L}{2 \pi } \mbox{Im} \log Z_1 \\ \nonumber
C_2 &=& -\frac{L^2}{2 \pi^2 } \mbox{Re} \log Z_1, 
\end{eqnarray}
where $\hat{X} = \sum_{j=1}^{N_e} \hat{x}_j$ denotes the total position operator.  From our derivations in the previous section it is obvious that higher order cumulants can also be obtained and that the finite difference approximation of the derivatives can also be improved, if desired.   \\

\subsection{Gauge invariant cumulants}

Souza, Wilkens, and Martin~\cite{Souza00} were the first to consider a generating function approach to the problem of crystalline polarization of band insulators.  In this approach the generating function takes the form,
\begin{equation}
\ln \mathcal{G}(\alpha) =\int_{-\frac{\pi}{l}}^{\frac{\pi}{l}} d k \langle u_k | u_{k+\alpha} \rangle,
\end{equation}
apart from a factor of $\frac{L}{2 \pi}$, which we removed here for convenience.  Cumulants and moments can be derived from $\mathcal{G}(\alpha)$ through,
\begin{eqnarray}
M_n^{(\mathcal{G})} &=& \frac{1}{i^n} \left. \frac{\partial^n \mathcal{G}(\alpha)}{\partial \alpha^n} \right|_{\alpha=0}, \\ \nonumber
C_n^{(\mathcal{G})} &=& \frac{1}{i^n} \left. \frac{\partial^n \ln \mathcal{G}(\alpha)}{\partial \alpha^n} \right|_{\alpha=0}.
\end{eqnarray}
We focus on the cumulants, of which the first two have the form,
\begin{eqnarray}
C_1^{(\mathcal{G})} &=& \left( \frac{ 1}{ i} \right)  \int_{-\frac{\pi}{l}}^{\frac{\pi}{l}} d k \langle u(k) | \frac{\partial}{\partial k} | u(k) \rangle = \gamma_Z, \\ \nonumber
C_2^{(\mathcal{G})} &=& \int_{-\frac{\pi}{l}}^{\frac{\pi}{l}} d k \left( - \langle u(k) | \frac{\partial^2}{\partial k^2} | u(k) \rangle + (\langle u(k) | \frac{\partial}{\partial k} | u(k) \rangle)^2 \right).
\end{eqnarray}
where $\gamma_Z$ is the Zak phase, obtained in Eq. (\ref{eqn:gamma_Z}).   Larger cumulants also take the forms of integrals over the Brillouin zone.  For insulators such cumulants correspond to $\mathcal{O}(1)$ numbers.\\

\subsection{Generating functions for the Berry phase}

In this subsection we state the quantity which serves as the generating function for the Berry phase and possible other higher cumulants.  This quantity is an extension of the Bargmann invariant.   We note that the connection between the Bargmann invariant and quantum geometry was investigated in two previous studies~\cite{Avdoshkin23,Avdoshkin25}.\\

The similarity between the discrete Berry phase, Eq. (\ref{eqn:g1_lnG1}), and the first cumulant in Eq. (\ref{eqn:C_periodic}) suggests that there may be a generating function from which the Berry phase itself, and higher order cumulants can be generated.  This is indeed the case.  The generating function in this case is an extended version of the Bargmann invariant, the cyclic product,
\begin{equation}
\label{eqn:Gamma_q}
\Gamma_q = \prod_{J=0}^{M-1} \langle \Psi_0 (\xi_J) | \Psi_0(\xi_{[J+q] \% M})\rangle,
\end{equation}
where the $[J+q]\%M$ denotes the remainder after division of $[J+q]$ by $M$.  As is the case with the original Bargmann invariant, $\Gamma_q$ is a {\it physically well-defined quantity} because the arbitrary phase of each $|\Psi(\xi_J)\rangle$ is canceled by its dual vector $\langle \Psi (\xi_J) |$.   In addition to the discrete Berry phase, it is possible to define moments and cumulants, the same way as in Eq. (\ref{eqn:MC_fd}),
\begin{eqnarray}
\label{eqn:MC_fd}
M_n^{(\Gamma)} &=& \left( \frac{1}{i} \right)^n \left. \frac{\delta^n}{ \delta q^n}  \Gamma_q \right|_{q=0}, \\ \nonumber
C_n^{(\Gamma)} &=& \left( \frac{1}{i} \right)^n \left. \frac{\delta^n}{ \delta q^n}  \mbox{Log} \Gamma_q \right|_{q=0},
\end{eqnarray}
\\

This is also possible for an open path Berry phase (Zak phase).  In this case one considers the parameter set, $\xi_0,...,\xi_{M+q-1}$, so that each of the last $q$ points is symmetry related to the first $q$ points, as
\begin{equation}
| \Psi (\xi_{M+J}) \rangle = \hat{S} | \Psi (\xi_J) \rangle.
\end{equation}
The extended Bargmann invariant, from which moments and cumulants can be derived, is
\begin{equation}
\label{eqn:Gamma_q_S}
\Gamma_q = \langle \Psi_0 (\xi_0) | \Psi_0(\xi_q)\rangle  \langle \Psi_0 (\xi_1) | \Psi_0(\xi_{1+q})\rangle...\langle \Psi_0 (\xi_{M-2}) | \Psi_0(\xi_{M-2+q})\rangle \langle \Psi_0 (\xi_{M-1}) | \Psi_0(\xi_{M-1+q})\rangle.
\end{equation}
The last $q$ terms in this product can be "pulled back" into the manifold of states $J=0,...,M-1$ by use of the symmetry operator $\hat{S}$.  Applying it to Eq. (\ref{eqn:Gamma_q_S}) results in,
\begin{eqnarray}
\label{eqn:Gamma_q_S2}
\Gamma_q = \langle \Psi_0 (\xi_0) | \Psi_0(\xi_q)\rangle  \langle \Psi_0 (\xi_1) | \Psi_0(\xi_{1+q})\rangle...\langle \Psi_0 (\xi_{M-2}) |\hat{S} |\Psi_0(\xi_{-2+q})\ \langle \Psi_0 (\xi_{M-1}) |\hat{S} |\Psi_0(\xi_{-1+q})\rangle.
\end{eqnarray}
$\Gamma_q$ is still independent of arbitrary phases because for each $| \Psi(\xi_J) \rangle$ the dual $\langle \Psi(\xi_J)|$ also appears in the product.  In the case, when the wave functions in question are Bloch functions, we can define the generating function as,
\begin{equation}
\label{eqn:Gamma_q_Zak}
\Gamma_q^{(Z)} = \prod_{m=0}^{N_c-1} \langle u_{k_m} | u_{k_{[m+q]\%N_c}} \rangle = Z_q.
\end{equation}
In some sense, the effect of the symmetry operator is the same as the modulo operation in the index of the ket vector of the scalar product.   In Eq. (\ref{eqn:Gamma_q_Zak}) we also indicate that $\Gamma_q^{(Z)}$ is equal to $Z_q$ when the wavefunction under scrutiny is the Slater determinant of a filled band.\\

As before, cumulants can be derived by applying the usual finite difference formulas according to,
\begin{eqnarray}
\label{eqn:MC_Z}
M_n^{(Z)} &=& \left( \frac{N_c l}{2 \pi i} \right)^n \left. \frac{\delta^n}{ \delta q^n}  Z_q \right|_{q=0}, \\ \nonumber
C_n^{(Z)} &=& \left( \frac{N_c l}{2 \pi i} \right)^n \left. \frac{\delta^n}{ \delta q^n}  \mbox{Log} Z_q \right|_{q=0},
\end{eqnarray}
The continuous Berry-Zak phase can be derived by taking the continuous limit (or the thermodynamic limit) of $C_1^{(Z)}$,
\begin{equation}
\lim_{N_c \rightarrow \infty} C_1^{(Z)} = \frac{L}{2 \pi }\gamma_Z = \frac{L}{2 \pi i} \int_{-\frac{\pi}{l}}^{\frac{\pi}{l}} d k \langle u_k | \frac{\partial}{ \partial k}| u_k \rangle.
\end{equation}
It will turn out to be useful to consider the limits of the quantities,
\begin{equation}
\label{eqn:c_Z}
c_n^{(Z)} = \frac{1}{i^n} \left. \frac{\delta^n}{ \delta q^n}  \mbox{Log} Z_q  \right|_{q=0}.
\end{equation}
For large $N_c$, $c_1^{(Z)}$ can be shown to be,
\begin{equation}
c_1^{(Z)} = \frac{\Delta k}{i} \sum_{m=0}^{N_c-1} \langle u_{k_m} | \frac{\partial}{\partial k} | u_{k_m} \rangle, 
\end{equation}
where $\Delta k = \frac{2 \pi}{N_c l}$.   $c_1^{(Z)}$ tends to $\gamma_Z$ in the thermodynamic limit.  Carrying out the same steps for $c_2^{(Z)}$ results in,
\begin{equation}
c_2^{(Z)} = (\Delta k)^2 \sum_{m=0}^{N_c} \left(-\langle u_{k_m} | \frac{\partial^2}{\partial k^2} | u_{k_m} \rangle + \left(\langle u_{k_m} | \frac{\partial}{\partial k} | u_{k_m} \rangle \right)^2 \right).
\end{equation}
The point to note is that $c_2^{(Z)}$ can not be turned into an integral with a finite value in the thermodynamic limit, because $\Delta k$ appears to the second power rendering the thermodynamic limit zero.   For any $n$ the reduced cumulant, $c_n^{(Z)}$ give expressions which are sums multiplied by $(\Delta k)^n$.   This means that when the continuous limit is taken, only $c_1^{(Z)}$ survives.    However, ratios of such cumulants, constructed according to the Binder criterion, can still give rise to a meaningful quantity, because in that case $(\Delta k)^n$ cancel.  \\

We now consider the excess kurtosis, or Binder cumulant, constructed from Eqs. (\ref{eqn:MC_Z}) in the thermodynamic limit in the insulating phase.  Using the relations established between the gauge invariant cumulants and $C_n^{(Z)}$ and $C_n^{(\mathcal{G})}$ it can be shown that,
\begin{equation}
\label{eqn:U4limit}
\lim_{N_c \rightarrow \infty} -\frac{1}{3} \frac{C_4^{(Z)}}{(C_2^{(Z)})^2} = \lim_{N_c \rightarrow \infty}-\frac{1}{3}\frac{2 \pi}{ N_c l} \frac{C_4^{(\mathcal{G})}}{(C_2^{(\mathcal{G})})^2} \rightarrow 0.
\end{equation}
This result depends crucially on the fact that the gauge invariant cumulants of Souza, Wilkens, and Martin~\cite{Souza00} give finite numbers for insulators.  Below, we will show that this is born out by numerical calculations.  

\subsection{Constructing the geometric Binder cumulant for quantum cycles}

The geometric Binder cumulant~\cite{Hetenyi19,Hetenyi22,Hetenyi25} (or excess kurtosis) in the context of adiabatic or quasiadiabatic cycles takes the form,
\begin{equation}
U_4 = 1 - \frac{M_4}{3 M_2^2},
\end{equation}
where $M_4$ and $M_2$ are to be obtained from $Z_q$ in the case of crystalline insulators.  A Binder cumulant can be defined for any Berry phase from the associated extended Bargmann invariant (Eq. (\ref{eqn:Gamma_q})).   \textcolor{black}{In numerical implementations }the Binder cumulant should be constructed from centered moments, rather than cumulants, because the cumulants in the case of discrete generating functions can become undefined due to the presence of terms such as $\mbox{Log} Z_q$~\cite{Hetenyi22}.   In insulators, this causes no difficulty, but as the polarization distribution flattens, $Z_q$ is expected to become sharply peaked around zero (in other words, it takes the form $Z_0 = 1, Z_q = 0, \forall q \neq0$).  In this case the logarithmic terms in the cumulants will diverge.  The flattening of the polarization is often, but not always, related to gap closure.  When a gap closes somewhere in the Brilloun zone, the scalar product $\langle u_{k_i} | u_{k_{i+1}} \rangle$ becomes zero because a level crossing occurred in the interval between $k_i$ and $k_{i+1}$.   The geometric phase becomes undefined in all of these cases, however,  the geometric Binder cumulant is well-defined and takes a finite value.  \textcolor{black}{While the choice of avoiding the logarithmic derivatives in the construction of the geometric Binder cumulant may appear arbitrary at this point, below, justification will be provided.  The geometric Binder cumulant will be calculated in several models, including a simple tight-binding model with open and periodic boundary conditions (section \ref{ssec:xmpls_FS}).   In subsection \ref{ssec:xmpls_AA} the second cumulant will be calculated based on the different approximation schemes.} \\

The excess kurtosis and the Binder cumulant correspond to centered distributions.  In most calculations the distribution is not directly available, it is the characteristic function, $Z_q$ which is.  One can effectively center the underlying distribution by taking the absolute value of each $Z_q$.   For the lowest order finite difference approximation $M_2$ and $M_4$ take the forms,
\begin{eqnarray}
\label{eqn:M2M4}
M_2 &=& \frac{L^2}{2 \pi^2} (1 - |Z_1|) \\ \nonumber
M_4 &=& \frac{L^4}{8 \pi^4} ( |Z_2| - 4 |Z_1| + 3). 
\end{eqnarray}
We will refer to this approach as the finite difference derivative (FDD) method.  Under these conditions, for the flat distribution, $U_4 = \frac{1}{2}$.   Since this is the lowest order finite difference approximation, we will refer to it as first order.  We will denote the order of approximation with the letter $m$.  The $\mu$ order approximation is a finite difference approximation accurate to order $\mathcal{O}(L^{-2\mu})$.  More accurate expressions can be obtained for moments and cumulants by applying higher order approximations to the finite difference derivatives.   The next order approximation to $M_2$ and $M_4$ (order $\mu=2$) gives,
\begin{eqnarray}
\label{eqn:M2M4_m2}
M_2 &=& \frac{L^2}{24 \pi^2} (|Z_2| - 16 |Z_1| + 15 ) \\ \nonumber
M_4 &=& \frac{L^4}{48 \pi^4} ( -|Z_3| + 12 |Z_2| - 39 |Z_1| + 28). 
\end{eqnarray}
In this case, the geometric Binder cumulant for the flat distribution changes slightly, to $U_4 = 0.50\bar{2}$.  

\section{Demonstrative examples}

\label{sec:xmpls}

In this section we validate the Binder cumulant as a localization probe for the metal-insulator transition.  We use two examples to do this: a simple Fermi sea (tight-binding model) and the Aubry-Andr\'{e} model.  The former models a simple conducting system with glosed gap, while the latter exhibits a metal-insulator transition, so it models both.  We will address the issue of increasing the order of the finite difference approximation. \\

\subsection{The Fermi sea}

\label{ssec:xmpls_FS}

The simple tight-binding model with open boundary conditions of length $L$ has a Hamiltonian of the form,
\begin{equation}
H = -t \sum_{j=1}^{L-1} \left( c_{j+1}^\dagger c_j + c_j^\dagger c_{j+1} \right). 
\end{equation}
We take the length of the unit cell is unity.  The single particle states are,
\begin{equation}
\phi_m(x_j) = \frac{1}{\sqrt{N_\phi}} \sin \left( \frac{m \pi x_j}{L+1} \right),
\end{equation}
where $m$ is a quantum number, $x_j=1,...,L-1$  and $N_\phi$ denotes a normalization constant.   The second and fourth moments of the position give,
\begin{eqnarray}
M_2 &=& \frac{1}{N}\sum_{m=1}^N \sum_{j=1}^L x_j^2|\phi_m(x_j)|^2 \\ \nonumber
M_4 &=& \frac{1}{N}\sum_{m=1}^N \sum_{j=1}^L x_j^4|\phi_m(x_j)|^2 
\end{eqnarray}
These are simple statistical moments of the distribution corresponding to $|\phi_m(x_j)|^2$. We calculated $M_2$, $M_4$ for a variety of system sizes at various particle densities.    The Binder cumulant tends to the value $0.4$ in the limit of large system size.  This is due to the fact that the distribution of the total position tends to a flat distribution, for which the Binder cumulant takes this known value.  \\

We now consider the same model under periodic boundary conditions.  There are two possibilities to consider, because the ground state, depending on the relative parity of particle number versus lattice size, is either non-degenerate (occurs either when both $N$ and $L$ are even, or both odd) of two-fold degenerate (even $N$, odd $L$, or vice versa).   To see how this state of affairs arises, we note that the quantity $Z_q$ is a scalar product, $\langle \Psi | \tilde{\Psi}\rangle$, where $|\tilde{\Psi}\rangle = \exp\left(i \frac{2\pi q}{L} \hat{X}\right)|\Psi\rangle$ denotes the ground state  with all momenta shifted by $\frac{2 \pi }{L} q$ as a result of the twist operator.  If the ground state is non-degenerate, then there is only one ground state, which has to have zero total momentum.  Since $Z_q$ will be the scalar product of a zero momentum state and one with a finite momentum (due to the action of $\hat{U}^q$), all $Z_q=0$, except for $Z_0=1$.  This gives a flat distribution, therefore, the expected value of geometric Binder cumulant is $U_4=0.4$.  \\

When the ground state is degenerate, the ground state wave function will have two components, one with total momentum $\pi/L$, the other with total momentum $-\pi/L$.  Let us write this ground state wave function as
\begin{equation}
|\Psi \rangle = a |\Psi_{\pi/L} \rangle + b |\Psi_{-\pi/L} \rangle,
\end{equation}
where $a$ and $b$ are two complex numbers, each with magnitude $1/\sqrt{2}$, because the total wave function has to have zero total momentum.  To calculate $Z_1$,  we apply the shift operator $\hat{U}$ once, resulting in
\begin{equation}
|\tilde{\Psi} \rangle = \hat{U} |\Psi \rangle =  a |\Psi_{3\pi/L} \rangle + b |\Psi_{\pi/L} \rangle.
\end{equation}
Evaluating the scalar product results in
\begin{equation}
Z_1 = \langle \Psi | \tilde{\Psi} \rangle = a^*b,
\end{equation}
from which it follows that $|Z_1| = 1/2$.  A similar analysis shows that other $Z_q=0$, except $Z_0=1$.   In this case, the underlying polarization distribution is the raised cosine distribution. \\

\begin{figure}[ht]
 \centering
 \includegraphics[width=12cm,keepaspectratio=true]{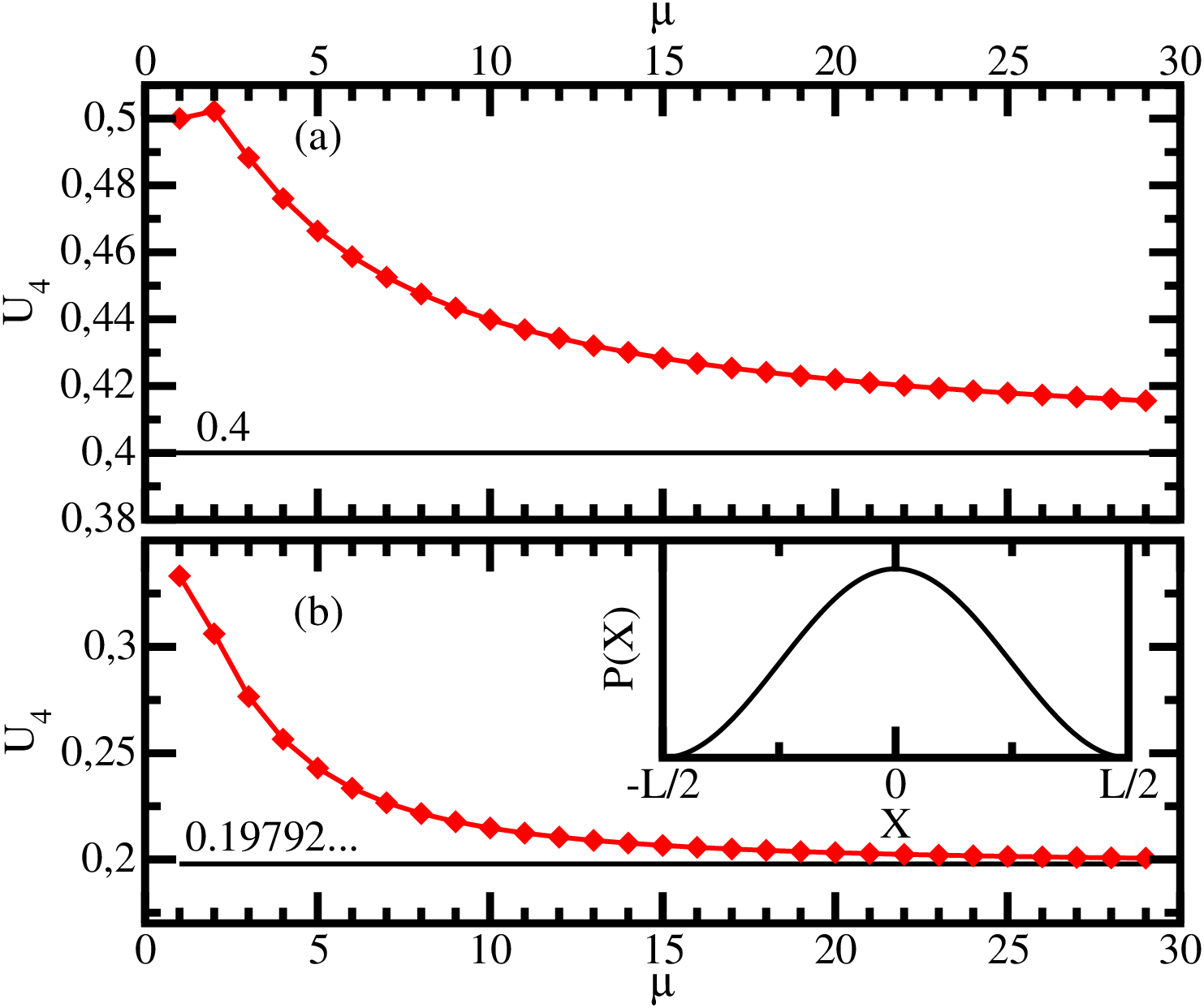}
 \caption{Geometric Binder cumulant ($U_4$) as a function of the order of the finite difference approximation ($\mu$), for the Fermi sea on a one-dimensional lattice.   Panel (a) shows results for the Fermi sea with non-degenerate ground state.  As $\mu$ is increased $U_4 \rightarrow 0.4$, which is the known result for a flat distribution.  Panel (b) shows results for the Fermi sea with two-fold degenerate ground state.  As $\mu$ is increased, $U_4 \rightarrow 0.19792...$, more precisely, minus one-third times the the number given in Eq. (\ref{eqn:RC}).  This is the known value of the Binder cumulant of a raised cosine distribution.  The inset shows the raised cosine distribution.  }
 \label{fig:U4_flat_rc}
\end{figure}

We calculated $U_4$ to higher order approximations and compared it to the expected values based on the known distributions.  Our results are shown in Fig. \ref{fig:U4_flat_rc}, part (a).  As the order of the finite difference approximation is improved, the calculated $U_4$ tends to $0.4$.  Part (b) of this figure shows the same calculations for the degenerate case.   As the finite difference approximation is improved, $U_4$ approaches its expected value,  minus one third times the constant given in Eq. (\ref{eqn:RC}).   The inset shows the raised cosine distribution.  \\

These results validate the use of the geometric Binder cumulant in the conducting phase.  Not only does it signal the delocalization associated with the transition, but it is also sensitive to the type of the underlying polarization distribution that is determined by degeneracy.   \\

\subsection{The Su-Schrieffer-Heeger model}

To demonstrate the applicability of the geometric Binder cumulant formalism to gap closure, we present calculations for the Su-Schrieffer-Heeger (SSH) model~\cite{Su79}.  This model was first derived in the context of analyzing solitons in one-dimensional organic conductors.  It is a tight-binding model of alternating hoppings, with Hamiltonian of the form,
\begin{equation}
H_{SSH} = \sum_{j=1}^L \left( J_o c_j^\dagger d_j + J_e d_j^\dagger c_{j+1} + \mbox{H. c.} \right).
\end{equation}
$J_e$ and $J_o$ denote alternating (odd vs. even) hopping strengths.  These parameters can also be written as,
\begin{equation}
J_o = \bar{J} + \delta J, \hspace{.5cm} J_e = \bar{J} - \delta J,
\end{equation}
where $\bar{J}$ denotes the average hopping strength, and $\delta J$ denotes the degree to which they alternate.  For $\delta J \neq 0$, this model is a gapped insulator. \\

\begin{figure}[ht]
 \centering
 \includegraphics[width=12cm,keepaspectratio=true]{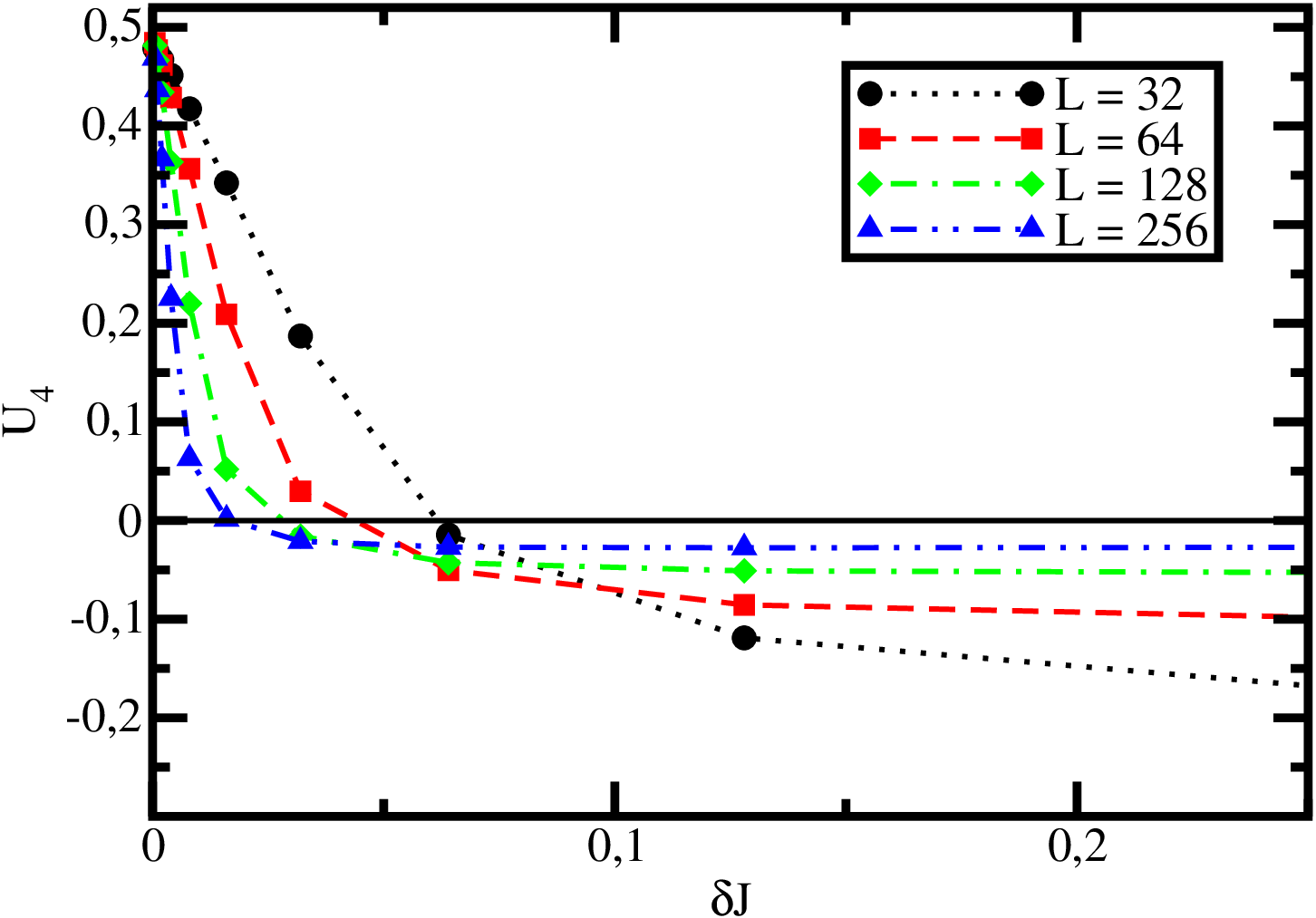}
 \caption{\textcolor{black}{Geometric Binder cumulant ($U_4$) as a function of the alternating part of the hopping parameter ($\delta J$) for the SSH model for various system sizes.  Gap closure occurs at $\delta J = 0$.   $U_4$ is calculated based on the finite difference approximation of order $\mathcal{O}(L^{-2})$.  The curves appear to converge to a value of $0$ for $\delta J \neq 0$ for the larger system sizes.  } }
 \label{fig:SSHU4}
\end{figure}

We calculated the geometric Binder cumulant, $U_4$, for SSH lattices with different system sizes as a function of $\delta J$.  Our results are shown in Fig. \ref{fig:SSHU4}.  At $\delta J = 0$, where gap closure occurs, $U_4 = 0.5$ for all system sizes.  As the gap is opened by increasing $\delta J$ $U_4$ decreases below zero.  For increasing system sizes the point at which $U_4 = 0$ is crossed approaches zero.  The curves appear to level of as $\delta J$ increases at some negative value.  The larger the system size the closer the value of $U_4$ is to zero.  This result is in accordance with Eq. (\ref{eqn:U4limit}).  \\

\subsection{Localization in the Aubry-Andr\'{e} model}

\label{ssec:xmpls_AA}

The Aubry-Andr\'{e} model was introduced~\cite{Aubry80} to capture the essence of what it means to be a quasicrystal.  The one-dimensional minimalist model consists of a hopping term and a potential term with a modulation  that depends on an irrational parameter.  The model and many of its generalizations have been studied~\cite{Martinez18,Billy08,Roati08,Dominguez-Castro19,Jitomirskaya99,Avila06,Avila09,Avila23,Modugno09,Wang17,Zhang15,Mastropietro15,Xu19,Cookmeyer20,Huang24,Varma15,Hetenyi24,Hetenyi25,Lu25,Goswami25,Zhang25} and still constitute an active research area.   Early studies focused on the single particle properties and it was shown by detailed mathematical analysis~\cite{Jitomirskaya99,Avila06,Avila09,Avila23} that they all become localized at finite interaction strength.  This state of affairs does not persist when the many particle, but still non-interacting, version of the model is considered.  There, due to the filling of quasiperiodic bands, the phase diagram of the model as a function of potential strength versus particle density corresponds to what mathematicians call an indicator function.  An example of an indicator function would be the Dirichlet function, a function which takes the value of zero for rational numbers, but a value of unity for irrational ones.  The phase diagram of the Aubry-Andr\'{e} model takes a finite value of the potential strength for rational fillings, but it goes to zero for fillings which approach a certain subset of irrational numbers in the large system limit.  The subset of irrational numbers depends on the irrational model parameter chosen.  Usually, the irrational parameter of the model is taken to be the inverse golden ratio.  In this case fillings which correspond to irrational numbers to which ratios of Fibonacci numbers (and sums of such numbers) tend in the large system limit exhibit no true phase transition, only a localized phase for any potential strength.  This is in contrast to rational fillings, for example, a filling of $N/L=1/2$, exhibits a localization transition at finite potential strength.   These results were found in Refs. ~\cite{Cookmeyer20,Hetenyi25} and will be explained further below.  \\

Due to the colorful features of the phase diagram, this model offers a great testing ground for the formalism considered here.   For the geometric Binder cumulant this has been done in Refs. \cite{Hetenyi24} and \cite{Hetenyi25}.    In this subsection we focus on the technical question of the type of finite difference approximation to be used.  We compare the Resta-Sorella expression for the variance, which is based on finite difference logarithmic derivatives, to the one based on simple finite difference derivatives (Eqs. (\ref{eqn:M2M4}) and (\ref{eqn:M2M4_m2})).  In the next subsection we will calcuate the fidelity susceptibilty for the Aubry-Andr\'{e} model.\\

The Hamtilonian of the Aubry-Andr\'{e} model is given by,
\begin{equation}
\label{eqn:HAA}
H = \sum_{j=0}^{L-1} -t(c_{j+1}^\dagger c_j + c_j^\dagger c_{j+1}) + W \cos(2 \pi \alpha j ) n_j,
\end{equation}
where $t$ denotes the hopping parameter and $W$ stands for the potential strength.  The model exhibits a metal-insulator transition if $\alpha$ is an irrational number.  As is most often done, we take $\alpha$ to be the golden ratio.  We assume periodic boundary conditions and approximate the irrational parameter as $\alpha = \frac{F_{n+1}}{F_n}$, the ratio of consecutive Fibonacci numbers.  We also take the system size to be $L = F_n$.  \\

In the calculation of the phase diagram of the AA model some technicalities are important.  When a many-body system is considered, the relative parity of the number of particles ($N$) and number of sites ($L$) has to be opposite for the polarization distribution to be a unimodal distribution.  It is in this case that the geometric Binder cumulant, $U_4$ can be applied.  For systems where $N$ and $L$ have the same parity, the distribution is bimodal. \\

\begin{figure}[ht]
 \centering
 \includegraphics[width=16cm,keepaspectratio=true]{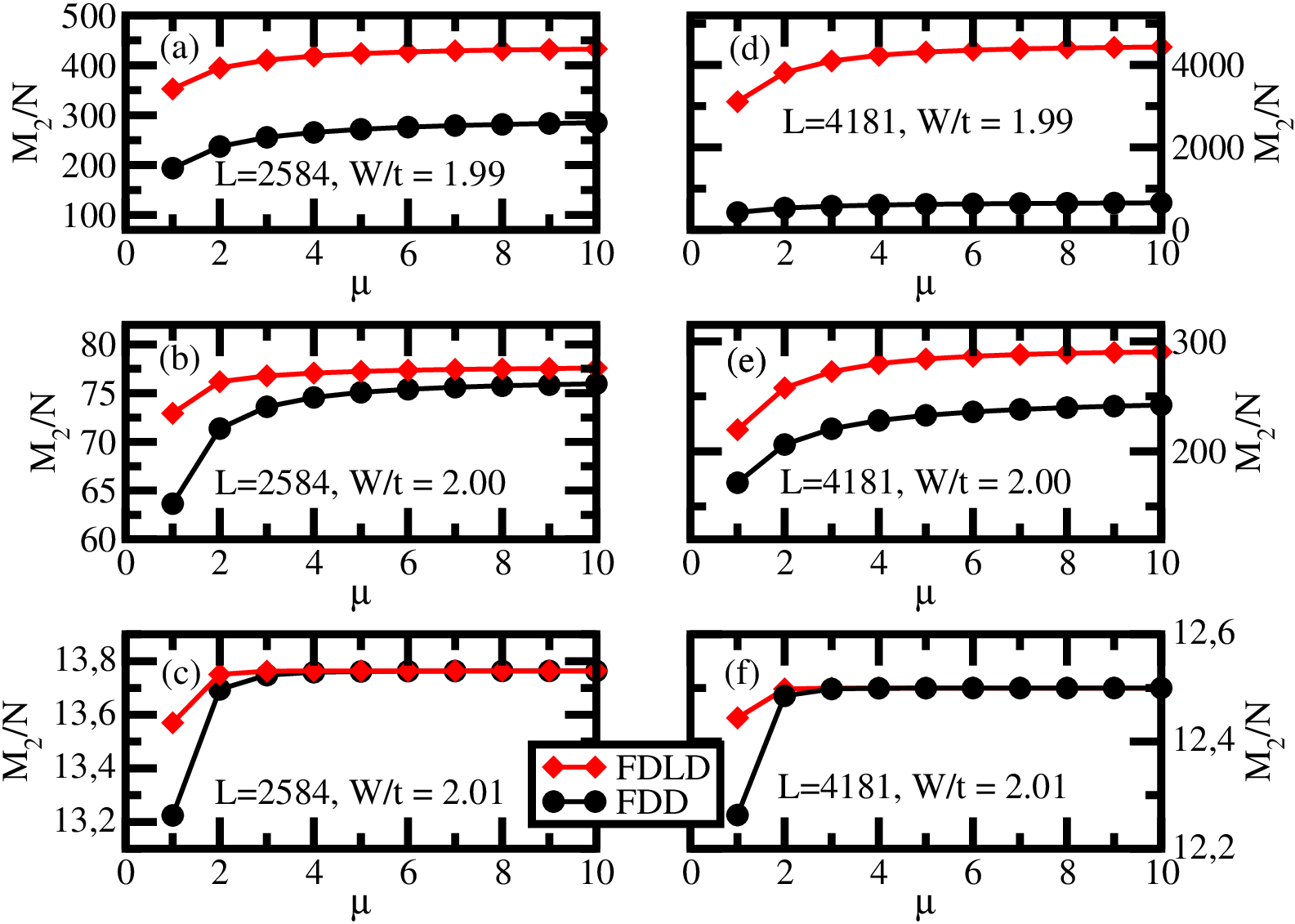}
 \caption{Comparison of different finite difference derivative schemes for the variance of the polarization divided by particle number.  FDLD refers to finite difference logarithmic derivative (the lowest order of which, $\mu=1$ is the Resta-Sorella scheme, given in Eq. (\ref{eqn:Resta_Sorella})), FDD refers to finite difference derivative (see Eq. (\ref{eqn:M2M4}) for the lowest order approximation, $\mu=1$, Eq. (\ref{eqn:M2M4_m2}), for the second lowest order approximation, $\mu=2$).  Shown are six calculations for the following parameter sets: (a) $L =2584,W/t=1.99$, (b) $L=2584, W/t = 2.00$, (c) $L=2584, W/t = 2.01$, (d) $L=4181, W/t = 1.99$, (e) $L=4181, W/t = 2.00$, and (f) $L = 4181, W/t = 2.01$.}
 \label{fig:L2.00}
\end{figure}

In our calculations we focused on what happens near the transition, $W=2t$.  In Fig. \ref{fig:L2.00} calculations are shown for two system sizes ($L=2584$ and $L=4181$) near half filling ($N=1291$ and $N=2090$), for three values of the potential strength, $W/t = 1.99, 2.00, 2.01$.   We present results for the second cumulant, calculated via Eqs. (\ref{eqn:M2M4}) and (\ref{eqn:M2M4_m2}) (as well as higher order finite difference approximations), as well as the Resta-Sorella scheme (and its higher order analogs).   The latter is referred to as the finite difference logarithmic derivative based approach.  The black filled circles stand for the former, while the red filled diamonds indicate the latter.   The axis indicates the order of the approximation, $\mu$.  For $W/t=1.99$ (delocalized or metallic phase) the two types of calculations differ greatly.  In the insulating phase ($W/t=2.01$) the two different types of calculations converge to the same value, although this does not happen at the level of approximation that is of widespread use.  $W/t=2.00$ is the critical point.  Calculations are published~\cite{Varma15,Hetenyi24} on this model at half-filling and it was found that the $M_2/N$ scales linearly with system size when the FDD scheme is applied, while the FDLD scheme does not lead to a simple power law scaling. \\

\subsection{The Aubry-Andr\'{e} transition through the fidelity suscebtibility}

The fidelity susceptibility can be viewed as the Provost-Vallee metric in a one dimensional parameter space.  We define it as,
\begin{equation}
\chi_F = \partial_\alpha \partial_{\alpha'} \left. \ln \langle \Psi(\alpha) | \Psi(\alpha') \rangle \right|_{\alpha' = \alpha}.
\end{equation}
Using the lowest order finite difference approximation leads to,
\begin{equation}
\chi_F = - \frac{1}{2 \delta^2} \mbox{Re} \ln \langle \Psi(\alpha - \delta) | \Psi (\alpha + \delta) \rangle,
\end{equation}
where $\delta$ is a small number.  \\

To understand the results we are about to present,  we will need a result from number theory, known as the Zeckendorf theorem~\cite{Zeckendorf72}.  This theorem states that all natural numbers can be written as a sum of Fibonacci numbers.  Most importantly, under certain constraints, this decomposition is unique.  Since $L$ is a Fibonacci number, any density can be written,
\begin{equation}
\label{eqn:Zeck}
\frac{N}{L} = \frac{\sum_{m=1}^M F_{I_m(N)}}{F_n}.
\end{equation}
In this expression, there are $M$ terms, also $M$ indices $I_m(N), m=1,...,M$, and they depend on $N$, the particle number, which was decomposed \`{a} la Zeckendorf.   Relevant to this study is the fact that any particle density (filling) can be written, uniquely, as a sum of ratios of Fibonacci numbers.   \\

Any ratio of Fibonacci numbers, $F_m/F_n$, approaches an irrational number, $\alpha(m,n)$, as the indices $m,n$ are shifted by the same integer, $s$, as follows,
\begin{equation}
\alpha(m,n) = \lim_{s \rightarrow \infty} \frac{F_{m+s}}{F_{n+s}}.
\end{equation}
Note that not all irrational numbers can be produced this way.  All irrational numbers of the form $\alpha(m,n)$ will depend on $\sqrt{5}$.   The larger $s$ is, the better $F_m/F_n$ approximates the irrational number $\alpha(m,n)$.  \\

Now consider all the fillings at some system size, $L=F_n$.  All densities, $N/L$ ,can be written according to the Zeckendorf decomposition (Eq. (\ref{eqn:Zeck})).  We can write an irrational number, $\tilde{\alpha}(N,L)$, corresponding to any density, $N/L$, as,
\begin{equation}
\tilde{\alpha}(N,L) = \lim_{s \rightarrow \infty} \frac{\sum_{m=1}^M F_{I_m(N)+s}}{F_{n+s}}.
\end{equation}
This allows extrapolation to irrational fillings by taking larger and larger system sizes through a systematic shift of $s$.   As an example, consider the filling, $F_{n-1}/F_n$.  As $n$ is increased, this number tends to $2/(1+\sqrt{5})=0.61803398875......$.   The worst approximation (if approximation is restricted to ratios of Fibonacci numbers) for this number is $F_1/F_2$, which is unity.  One filling we will study is $N/L = 377/610$, which is $F_{14}/F_{15}=0.61803278688...$, which is in agreement with its infinite limit irrational number up to $5$ significant digits.   The number $F_{14}/F_{15}$ is a {\it good} approximation to the number,
\begin{equation}
\tilde{\alpha}(377,610) = \lim_{s \rightarrow} \frac{F_{14+s}}{F_{15+s}}.
\end{equation}
We will contrast this case  with the filling $N/L = 379/610$.   The Zeckendorf decomposition in this case gives,
\begin{equation}
\frac{N}{L} = \frac{377 + 2}{610} = \frac{F_{14}}{F_{15}} + \frac{F_3}{F_{15}}.
\end{equation}
The first term of this sum is the same as the previous filling, already determined to be a good approximation to its own irrational limit.  But what about the second term?  The second term gives $2/610 = 0.00327868852...$, which tends to an irrational number, $0.00310563145..$, which can be obtained by taking the ratio of shifted indices (I did it by calculating $F_{13}/F_{25}$).  Agreement in this case is only obtained up to one significant figure.  It follows that $N/L=379/610$ is a {\it bad} approximation to the irrational number for
\begin{equation}
\tilde{\alpha}(379,610) = \lim_{s \rightarrow \infty} \left( \frac{F_{14+s}}{F_{15+s}} + \frac{F_{3+s}}{F_{15+s}} \right).
\end{equation}
A number $F_m/F_n$ tends to be a bad approximation if $m$ or $n$ is a small number.  In a particle density, $N/L<1$, so bad approximations are number such as $F_1/L$, $F_2/L$, $F_3/L$.   When added to numbers which are already good approximations, the overall filling turns into a bad approximation.\\

\begin{figure}[ht]
 \centering
 \includegraphics[width=14cm,keepaspectratio=true]{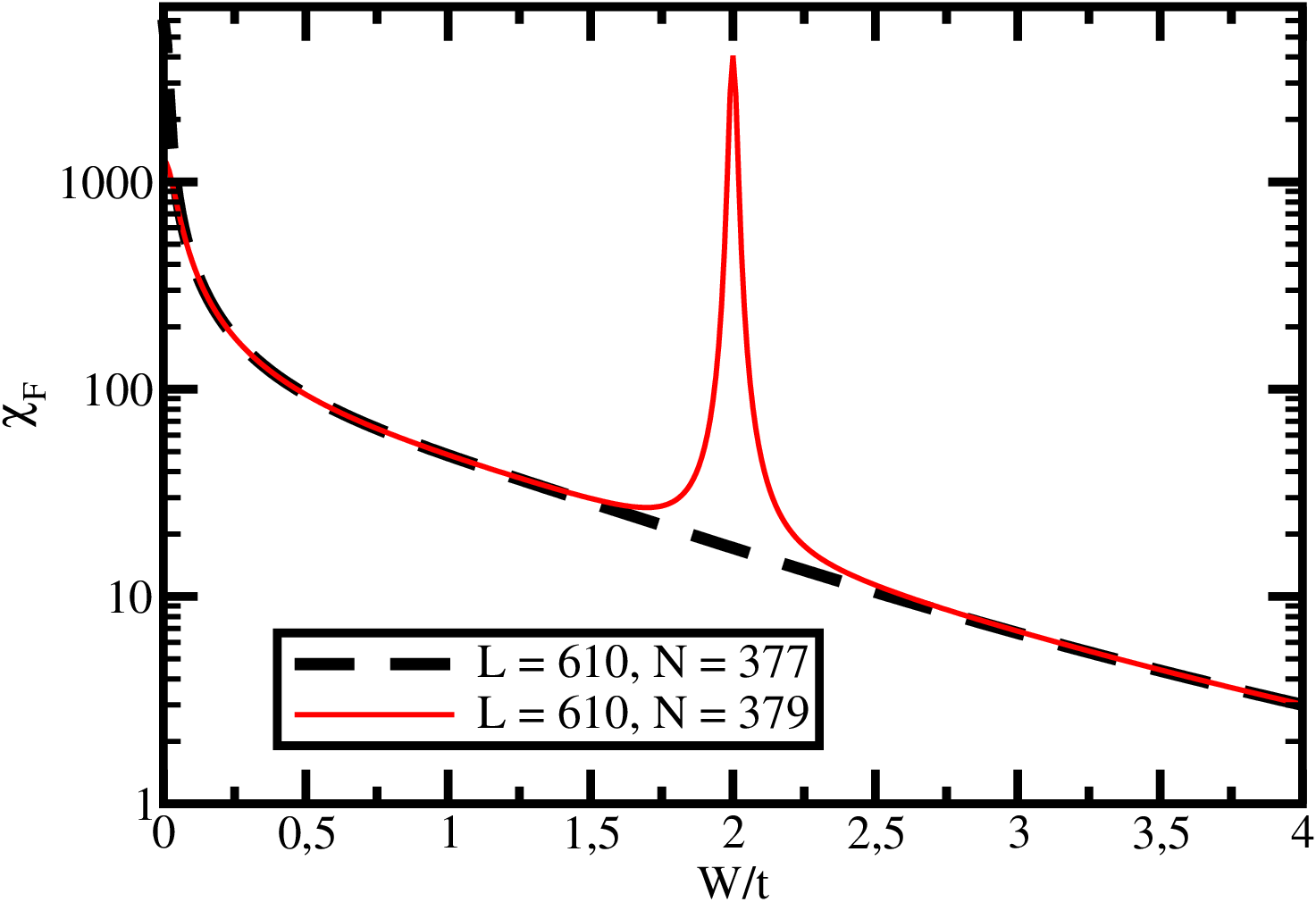}
 \caption{Fidelity susceptibility, $\chi_F$, for an Aubry-And\'{e} one-dimensional lattice as a function of $W/t$ for a system of size $L=610$ for two different particle numbers, $N=377$ and $N=379$.  The system with $N=377$ shows a singularity at $W/t=0$, the one with $N=379$ shows two singularities, one at $W/t=0$ and one at $W/t=2$.  The small parameter, defining the finite difference derivative, is $\delta = 0.01$.  The different behaviors for the two system sizes can be understood based on the Zeckendorf decomposition of natural numbers.  For the full explanation see the text.}
 \label{fig:chiF}
\end{figure}

We calculated the fidelity susceptibility ($\chi_F$) as a function of $W/t$ for the Aubry-Andr\'{e} model for fillings $N/L=377/610$ and $N/L=379/610$.  Our results are shown in Fig. \ref{fig:chiF}.   The small finite difference parameter is given by $\delta=0.01$.  Calculations for two particle numbers are shown, $N=377$ and $N=379$.   The "good" approximation, $N/L=377/610$ shows only one peak at $W/t=0$, meaning that the model exhibits localization for finite  $W/t$.   At $W/t=0$ gap closure occurs, the system becomes a simple Fermi sea.  The "bad" approximation, $N/L=379/610$ shows two peaks in $\chi_F$, one at $W/t=0$, the other at $W/t=2$, where the known localization transition in single transition states occurs.  Exactly these results were obtained in Refs. \cite{Cookmeyer20,Hetenyi25}. \\

Consider what happens when the system size is increased from $L=F_n$ to $L=F_{n+1}$.  All fillings that were present for size $L=F_n$ ($N = 1,...,F_n$) will be present at $F_{n+1}$, since all $N$ can be deomposed via the Zeckendorf decomposition.  These fillings will be improved approximations to their limit $\tilde{\alpha}(N,L)$.  But for the new system size, $L=F_{n+1}$ there will be new particle numbers, which are not good approximations (since they differ from the ones that were present for $L=F_n$ by small numbers).  For this reason, for any system size $L$ there will be particle densities for which all the model is localized for all finite $W/t$, as well as particle densities for which the delocalization-localization transition occurs at $W/t=2$.\\

For completeness, let us discuss the case of rational fillings.   For further results, see Refs. \cite{Varma15,Hetenyi24,Hetenyi25} .  The question is, how does one represent a rational number in the thermodynamic limit by a Zeckendorf decomposition.  A rational filling will require larger and larger Zeckendorf sums (more terms in Eq. (\ref{eqn:Zeck})), because, as the thermodynamic limit is approached, a rational number is being approximated by sum of irrational numbers.  This is interesting, because usually the question in number theory is how to approximate an irrational number as a sum of rational numbers.  To give a sense of this, let us consider a filling of $N/L=1/2$.   We will consider three system sizes, $L = 34, 144, 610$, corresponding to particle numbers of $N = 17, 72, 305$, respectively.  The Zeckendorf decompositions for these particle numbers are,
\begin{equation}
N = 13 + 3 + 1; N = 55 + 13 + 3 + 1; N = 233 + 55 + 13 + 3 + 1.
\end{equation}  
Most importantly, the number one is always present, so all calculations will result in a delocalization-localization transition at $W/t=2$.  Another key feature is that as the thermodynamic limit is taken the Zeckendorf sum becomes {\it infinite}, which was not the case for the category of irrational numbers already discussed.  As for irrational numbers which can not be produced as a sum of ratios of Fibonacci numbers via a Zeckendorf decomposition of $N$, we conjecture that $W/t=2$, because one has to approximate them in terms of Fibonacci ratios when the thermodynamic limit is taken.\\

Also, our numerical results for the phase transition indicate that the fidelity susceptibility, like the geometric Binder cumulant, has a practical advantage over the variance ($M_2$ or the Resta-Sorella coherence length of Ref. \cite{Resta99}).  To locate the phase transition point accurately, it is not necessary to simulate different system sizes, one system size suffices.  The "divergence", indicating the transition, is already visible for a single system size (Fig. \ref{fig:chiF}). \\

\section{Conclusion}

\label{sec:cnclsn}

In this work the geometric Binder cumulant construction was reviewed.  Two questions were posed in the introduction.  One was, whether it is possible to extend the geometric phase formalism to the case of quasiadiabatic cycles, that is, cycles which cross degeneracy points.  The second one was whether it is possible to construct a Binder cumulant in cases in which the fundamental quantity is a geometric phase, rather than an operator expectation value.  By answering both of these questions in the affirmative and explicitly constructing the relevant cumulants, it became possible to introduce a formalism in the spirit of the Binder cumulant in the modern theory of polarization.  The formalism was validated through example calculations.  The simplest one of these was the Fermi sea, a well known conductor.  This example was crucial, since the modern polarization theory is already well established for insulators, but its cumulants diverge in conductors.  The formalism was further validated by calculations in the Su-Schrieffer-Heeger and Aubry-Andr\'{e} models.   A complementary calculation using the fidellity susceptibility was also presented.  In all our calculations the essential element was the construction of the generating function, which turned out to be an extension of the Bargmann invariant.   Also essential, when it comes to implementation, is the type of finite difference derivative approximation used.  We showed that a centered moment based approximation which does not involve directly taking finite difference derivatives of the logarithm of the discrete characteristic function is most efficient, it preserves the size scaling on both sides of the metal-insulator transition.\\

Future directions may include extending the ideas presented here to topological insulators.  In topological insulators the relevant quantity, the topological invariant, is also a geometric phase, so, at least in principle, there appears to be no obstacle in constructing generating functions based on extended Bargmann invariants.  This also appears to be the case for the degenerate extensions of the geometric phase where the generating function is a Wilson loop.\\

\section*{Acknowledgements}
This research was supported by HUN-REN 3410107 (HUN-REN-BME-BCE Quantum Technology Research Group), by the National Research, Development and Innovation Fund of Hungary within the Quantum Technology National Excellence Program (Project No. 2017-1.2.1-NKP-2017-00001), by Grants No. K142179, No. K142652, and No. FK142601 and by the BME-Nanotechnology FIKP Grant No. (BME FIKP-NAT).


\begin{thebibliography}{9}

\bibitem{Lukacs60} E. Lukacs, "Characteristic Functions", Charles Griffin and Company, London, (1960).

\bibitem{Kubo62} R. Kubo, "Generalized Cumulant Expansion Method" {\it J. Phys. Soc. Japan} {\bf 17} 1100 (1962).

\bibitem{Fulde95} P. Fulde, "Electron Correlation in Molecules and Solids", Springer, Berlin, (1995).

\bibitem{Pancharatnam56} S. Pancharatnam, "Generalized Theory of Interference, and Its Applications. Part I. Coherent Pencils", {\it Proc. Indian Acad. Sci. A}
{\bf 44} 247 (1956).

\bibitem{LonguetHiggins58} H. C. Longuet-Higgins, U. \"{O}pik, M. H. L. Pryce, and  R. A. Sack, "Studies of the Jahn-Teller effect .II. The dynamical problem". Proc. R. Soc. A. 244 1 (1958).

\bibitem{Berry84} M. V. Berry, "Quantal Phase Factors Accompanying Adiabatic Changes", {\it Proc. Roy. Soc. A} {\bf 392} 45 (1984).

\bibitem{Wilczek89} F. Wilczek and A. Shapere, A., eds. "Geometric Phases in Physics", World Scientific, Singapore (1989).

\bibitem{Xiao10} D. Xiao, M.-C. Chang, and Q. Niu, "Berry phase effects on electronic properties", {\it Rev. Mod. Phys.} {\bf 82} 1959 (2010).

\bibitem{Kato50} T. Kato, "On the Adiabatic Theorem of Quantum Mechanics", {\it J. Phys. Soc. Jpn.} {\bf 5} 435 (1950).

\bibitem{Zak89} J. Zak, "Berry’s phase for energy bands in solids", {\it Phys. Rev. Lett.} {\bf 62} 2747 (1989).

\bibitem{King-Smith93} R. D. King-Smith and D. Vanderbilt, "Theory of polarization in crystalline solids", {\it
  Phys. Rev. B} {\bf 47} R1651 (1993).

\bibitem{Resta94} R. Resta, "Macroscopic polarization in crystalline dielectrics: the geometric phase approach",  {\it Rev. Mod. Phys.} {\bf 66} 899
  (1994).
  
\bibitem{Resta98} R. Resta, "Quantum Mechanical Position Operator in Extended Systems", {\it Phys. Rev. Lett.} {\bf 80} 1800 (1998).

\bibitem{Resta99} R. Resta and S. Sorella, "Electron Localization in the Insulating State", {\it Phys. Rev. Lett.} {\bf 82} 370 (1999).

\bibitem{Aligia99} A. A. Aligia and G. Ortiz, "Quantum Mechanical Position Operator and Localization in Extended Systems", {\it Phys. Rev. Lett.} {\bf 82} 2560 (1999).

\bibitem{Ortiz00} G. Ortiz and A. A. Aligia, "How Localized is an Extended Quantum System?", {\it Phys. Stat. Solidi B} {\bf 220} 737 (2000).

\bibitem{Souza00} I. Souza, T. Wilkens, and R. M. Martin, "Polarization and localization in insulators: Generating function approach",{\it
  Phys. Rev. B} {\bf 62} 1666 (2000).

 \bibitem{Resta00} R. Resta, "The insulating state of matter: a geometrical theory",  {\it J. Phys.: Cond. Mat.} {\bf 12} R107
  (2000).
  
\bibitem{Resta07} R. Resta and D. Vanderbilt, "Theory of polarization: a modern approach", in "Physics of Ferroelectrics: A Modern Perspective", eds. K. M. Raabe, C.-H. Ahn, and J.-M. Triscone, Springer Series in Topics in Applied Physics, vol. 105, Spriger Berlin Heidelberg (2007).
  
\bibitem{Resta10} R. Resta, "Electrical polarization and orbital magnetization: the modern theories", {\it J. Phys.: Cond. Mat.} {\bf 22} 123201 (2012).
  
\bibitem{Spaldin12} N. A. Spaldin, "A beginner's guide to the modern theory of polarization" {\it J. Solid State Chem.} {\bf 195} 2 (2012).
  
\bibitem{Vanderbilt18} D. Vanderbilt, {\it Berry Phases in Electronic Structure Theory}, Cambridge University Press, Cambridge, U.K. (2018).

\bibitem{Grosso00} G. Grosso and G. Pastori Parravicini, "Solid State Physics", {\it Academic Press}, London (2000).
  
\bibitem{Resta24} R. Resta, "Berry phase and geometrical observables", Elsevier, (2024).

\bibitem{Cardy96} J. L. Cardy, {\it Scaling and Renormalization in Statistical Physics}, Cambridge Lecture Notes in Physics, Cambridge University Press (1996).

\bibitem{Binder81a} K. Binder, "Finite size scaling analysis of ising model block distribution functions", {\it Z. Phys. B } {\bf 43} 119 (1981).
  
\bibitem{Binder81b} K. Binder, "Critical Properties from Monte Carlo Coarse Graining and Renormalization", {\it Phys. Rev. Lett. } {\bf 47} 693 (1981).

\bibitem{Fisher72a} M.~E. Fisher, in {\it Critical Phenomena},
  Proc. 51st Enrico Fermi Summer School, Varena, edited by M. S. Green
  (Academic Press, N.Y.) 1972.
  
\bibitem{Fisher72b} M.~E. Fisher and M.~N. Barber, "Scaling Theory for Finite-Size Effects in the Critical Region",{\it
  Phys. Rev. Lett.} {\bf 28} 1516 (1972).

\bibitem{Kohn64} W. Kohn, "Theory of the Insulating State", {\it Phys. Rev.} {\bf 133} A171 (1964).
  
\bibitem{Bernevig13} B. A. Bernevig and T. L. Hughes, {\it Topological Insulators and Superconductors}, Princeton University Press (2013).

\bibitem{Asboth16} J. K. Asb\'{o}th, L. Oroszl\'{a}ny, and A. P\'{a}lyi, {\it A Short Course on Topological Insulators: Band Structure and Edge States in One and Two Dimensions}, Lecture Notes on Physica, vol. 919, Springer International Publishing, (2016).

\bibitem{Qi11} X.-L. Qi and S.-C. Zhang, "Topological insulators and superconductors", {\it Reviews of Modern Physics}, {\bf 83} 1057 (2011).

\bibitem{Sato17} M. Sato and Y. Ando, "Topological superconductors: a review", {\it Rep. Prog. Phys.} {\bf 80} 076501 (2017).

\bibitem{Torma23} P. T\"{o}rm\"{a}, "Essay: Where Can Quantum Geometry Lead Us?" {\it Phys. Rev. Lett.} {\bf 131} 240001 (2023).

\bibitem{Provost80} J. Provost and G. Vallee, “Riemannian structure on manifolds of quantum states,” {\it Comm. Math. Phys.} {\bf 76}, 289 (1980).

\bibitem{Hetenyi23} B. Het\'{e}nyi and P. L\'{e}vay, "Fluctuations, uncertainty relations, and the geometry of quantum state manifolds", {\it Phys. Rev. A} {\bf 108} 032218 (2023).

\bibitem{Peotta15} S. Peotta and P. T\"{o}rm\"{a}, "Superfluidity in topologically nontrivial flat bands" , {\it Nat. Comm.} {\bf 6} 8944 (2015).

\bibitem{Yu25} J. Yu, B. A. Bernevig, R. Queiroz, E. Rossi, P. T\"{o}rm\"{a} and B.-J. Yang, "Quantum geometry in quantum materials" {\it npj Quantum Mater.} {\bf 10} 101 (2025).

\bibitem{Thompson25} J. J. P. Thompson, W. J. Jankowski, R.-J. Slager, B. Monserrat,  "Topologically enhanced exciton transport" {\it  Nat Commun} {\bf 16}, 11448 (2025).

\bibitem{Xie25} Y. Xie, R. Liu, and B. Gu, "Quantum geometrical molecular dynamics", {\it Sci. Adv.} {\bf 11} (2025).

\bibitem{Verma25} N. Verma and R. Queiroz, "Instantaneous response and quantum geometry of insulators" {\it Proc. Nat. Acad. Sci. USA} {\bf 122} e2405837122 (2025).

\bibitem{Pellitteri25} G. Pellitteri, Z. Dai, H. Hu, Y. Jiang, G. Menichetti, A. Tomadin,  B. A. Bernevig, and M. Polini, "Phonon spectra, quantum geometry, and the Goldstone theorem", {\it Phys. Rev. B} {\bf 112} 245128 (2025).

\bibitem{Avdoshkin23}A .Avdoshkin and F. K. Popov, "Extrinsic geometry of quantum states" {\it Phys. Rev. B} {\bf 107} 245136 (2023).

\bibitem{Avdoshkin25} A. Avdoshkin, J. Mitscherling, J. E. Moore, "Multistate geometry of shift current and polarization" {\it Phys. Rev. Lett.} {\bf 135} 066901 (2025).

\bibitem{Gu10} S.-J. Gu, "Fidelity Approach to Quantum Phase Transitions", {\it Int. J. Mod. Phys. B} {\bf 24} 4371 (2010).

\bibitem{Sachdev11} S. Sachdev, "Quantum Phase Transitions", 2nd ed., Cambridge University Press, Cambridge, UK, (2011).
  
\bibitem{Bargmann64} V. Bargmann, "Note on Wigner's Theorem on Symmetry Operations ", {\it J. Math. Phys.} {\bf 5} 862 (1964).

\bibitem{Hetenyi19} B. Het\'{e}nyi and B. D\'{o}ra, "Quantum phase transitions from analysis of the polarization amplitude", {\it Phys. Rev. B} {\bf 99} 085126 (2019).

\bibitem{Hetenyi22} B. Het\'{e}nyi and S. Cengiz, "Geometric cumulants associated with adiabatic cycles crossing degeneracy points: Application to finite size scaling of metal-insulator transitions in crystalline electronic systems", {\it Phys. Rev. B} {\bf 106}  195151 (2022).

\bibitem{Su79}W. P. Su, J. R. Schrieffer, and A. J. Heeger, "Solitons in polyacetylene", {\it Phys. Rev. Lett.} {\bf 42}, 1698 (1979).

\bibitem{Aubry80} S. Aubry and G. Andr\'{e}, "Analyticity breaking and Anderson localization in incommensurate lattices", {\it Ann. Isr. Phys.} {\bf 3} 133 (1980).

\bibitem{Martinez18} A. J. Martinez, M. A. Porter, and P. T. Kevrekidis, "Quasiperiodic granular chains and Hofstadter butterflies", {\it Philos. Trans. A} {\bf 376} 20170139 (2018). 

\bibitem{Billy08} J. Billy, V. Josse, Z. Zuo, A. Bernard, B. Hambrecht, P. Logan, D. Cl\'{e}ment, L. Sanchez-Palencia, P. Bouyer, and A. Aspect, "
Direct observation of Anderson localization of matter waves in a controlled disorder", {\it Nature (London)} {\bf 453} 891 (2008).

\bibitem{Roati08} G. Roati, C. D'Errico, L. Fallani, M. Fattori, C. Fort, M. Zaccanti, G. Modugno, M. Modugno, and M. Inguscio, "Anderson localization of a non-interacting Bose–Einstein condensate", {\it Nature (London)} {\bf 453} 895 (2008).

\bibitem{Dominguez-Castro19} G. A. Dominguez-Castro, R. Paredes, "The Aubry–André model as a hobbyhorse for understanding the localization phenomenon", {\it Eur. J. Phys.} {\bf 40} 045403 (2019).

\bibitem{Jitomirskaya99} S. Ya. Jitomirskaya, "Metal-insulator transition for the almost Mathieu operator", {\it Ann. Math.} {\bf 150}  1159 (1999).

\bibitem{Avila06} A. Avila, S. Jitomirskaya, "Solving the Ten Martini Problem"  In: J. Asch, A. Joye,  (eds) {\it Mathematical Physics of Quantum Mechanics}, Lecture Notes in Physics, vol 690. Springer, Berlin, Heidelberg . 

\bibitem{Avila09} A. Avila and S. Jitomirskaya "The Ten Martini Problem" {\it Ann. Math.} {\bf 170} 303 (2009).

\bibitem{Avila23} A. Avila, J. You, and Q. Zhou, "Dry Ten Martini Problem in the non-critical case" arxiv:2306.16254. 
 
 \bibitem{Modugno09} M. Modugno, "Exponential localization in one-dimensional quasi-periodic optical lattices", {\it New. J. Phys.} {\bf 11} 033023 (2009).
 
\bibitem{Wang17} Y. Wang, G. Xianlong, and S. Chen, "Almost mobility edges and the existence of critical regions in one-dimensional quasiperiodic lattices", {\it Eur. Phys. J. B} {\bf 90} 215 (2017).

\bibitem{Zhang15} Y. Zhang, D. Bulmash, A. V.  Maharaj, C.-M. Jian, and S. A. Kivelson, "The almost mobility edge in the almost Mathieu equation
", arXiv:1504.05205.

\bibitem{Mastropietro15} V. Mastropietro, "Localization of Interacting Fermions in the Aubry-André Model" {\it Phys. Rev. Lett.} {\bf 115} 180401 (2015).

 \bibitem{Xu19} S. Xu, X. Li, Y.-T. Hsu, B. Swingle, S. Das Sarma, "Butterfly effect in interacting Aubry-Andre model: Thermalization, slow scrambling, and many-body localization", {\it Phys. Rev. Research } {\bf 1}  032039(R) (2019).
 
 \bibitem{Cookmeyer20} T. Cookmeyer, J. Motruk, J. E. Moore, "Critical properties of the ground-state localization-delocalization transition in the many-particle Aubry-André model", {\it Phys. Rev. B} {\bf 101}174203 (2020).
 
 \bibitem{Huang24} K. Huang, D. Vu, S. Das Sarma, X. Li, "Interaction-enhanced many body localization in a generalized Aubry-Andre model" {\it Phys. Rev. Res.} {\bf 6}, L022054 (2024).
 
 \bibitem{Varma15} V. Kerala Varma and S. Pilati, "Kohn's localization in disordered fermionic systems with and without interactions" {\it Phys. Rev. B} {\bf 92} 134207 (2015).
 
 \bibitem{Hetenyi24} B. Het\'{e}nyi, "Scaling of the bulk polarization in extended and localized phases of a quasiperiodic model", {\it Phys. Rev. B} {\bf 110} 125124 (2024).
 
 \bibitem{Hetenyi25} B. Het\'{e}nyi and I. Balogh, "Numerical study of the localization transition of Aubry-André type models" {\it Phys. Rev. B} {\bf 112} 144203 (2025).
 
 \bibitem{Lu25} F. Lu, A. Zhou, S. Cheng, and G. Xianlong, "Wigner distribution, Wigner entropy, and anomalous transport of a generalized Aubry-André model" {\it Phys. Rev. B} {\bf 112} 174206 (2025).

\bibitem{Goswami25} A. Goswami, P. Chatterjee, R. Modak, and S. Sahoo, "Subsystem localization in a two-leg ladder system" {\it Phys. Rev. B} {\bf 112} 144205 (2025).

\bibitem{Zhang25} Z.-H. Zhang, H.-C. Kou, and P. Li, "Critical dynamics and its interferometry in the one-dimensional $p$-wave-paired Aubry-André-Harper model", {\it Phys. Rev. B} {\bf 112} 014310 (2025).

\bibitem{Zeckendorf72} E. Zeckendorf, ”Repr\`{e}sentation des nombres naturels par
une somme de nombres de Fibonacci ou de nombres de
Lucas”, Bull. Soc. R. Sci. Li\`{e}ge {\bf 41} 179 (1972).



\end{thebibliography}
\end{document}